\documentclass[submit]{aiaa-pretty}  

\usepackage{mathtools}    % need for sub equations
\usepackage{amsfonts}
\usepackage{graphicx}   % need for figures
\usepackage{subcaption}
\usepackage{epsfig} 
\usepackage{cancel}
\usepackage{amssymb}
\usepackage{color}
\usepackage{bm}
\usepackage[ruled,vlined,titlenotnumbered]{algorithm2e} 
\usepackage{todonotes} 
\usepackage{float}
\usepackage{cite}
\usepackage{enumitem}

\newcommand{\R}{\mathbb{R}} % Real number
\newcommand{\rc}{R_c} % Capture radius

\newcommand{\N}{N} % number of agents

\newcommand{\veh}{Q} % vehicle
\newcommand{\state}{x} % state
\newcommand{\ctrl}{u} % control
\newcommand{\dstb}{d} % disturbance
\newcommand{\pos}{p} % position
\newcommand{\npos}{h} % non-position states

\newcommand{\fdyn}{f} % full dynamics
\newcommand{\cset}{\mathcal{U}} % Control set
\newcommand{\cfset}{\mathbb{U}} % control function set
\newcommand{\dset}{\mathcal{D}} % disturbance
\newcommand{\dfset}{\mathbb{D}} % disturbance function set
\newcommand{\obsset}{\mathcal{G}} % Obstacle (the one used to solve PDE)
\newcommand{\dz}{\mathcal{Z}} % danger zone
\newcommand{\valfunc}{V} % value function
\newcommand{\brs}{\mathcal{V}} % backwards reachable set
\newcommand{\pfrs}{\mathcal{P}} % projected forwards reachable set
\newcommand{\targetset}{\mathcal{L}} % target set
\newcommand{\ham}{H} % Hamiltonian
\newcommand{\fc}{l} % Final condition
\newcommand{\ic}{l} % Initial condition
\newcommand{\obsfunc}{g} % Obstacle function
\newcommand{\costate}{\lambda}

\newcommand{\disckernel}{\Omega} % Discriminating kernel

\newcommand{\ldt}{t^\text{LDT}} % latest departure time
\newcommand{\sta}{t^\text{STA}} % scheduled time of arrival
\newcommand{\ioset}{\mathcal{O}} % Induced obstacle
\newcommand{\boset}{\mathcal{M}} % Base obstacle
\newcommand{\soset}{\ioset^\text{static}} % static obstacle in state space
\newcommand{\errorbound}{\mathcal{E}} % Error ``bubble" between vehicle and tracking reference
\newcommand{\tracklaw}{\kappa} % Robust tracking law

\newtheorem{alg}{Algorithm}

\title{\LARGE \bf Provably Safe and Robust Drone Routing via Sequential Path Planning: A Case Study in San Francisco and the Bay Area}

\author{Mo Chen\thanks{PhD Candidate, Department of Electrical Engineering and Computer Sciences, shared first author}, Somil Bansal\thanks{PhD Student, Department of Electrical Engineering and Computer Sciences, shared first author}, Ken Tanabe\thanks{Toshiba}, and Claire J. Tomlin\thanks{Professor, Department of Electrical Engineering and Computer Sciences, Member AIAA}
}

\AIAAabstract{Provably safe and scalable multi-vehicle path planning is an important and urgent problem due to the expected increase of automation in civilian airspace in the near future. Hamilton-Jacobi (HJ) reachability is an ideal tool for analyzing such safety-critical systems and has been successfully applied to several small-scale problems. However, a direct application of HJ reachability to large scale systems is often intractable because of its exponentially-scaling computation complexity with respect to system dimension, also known as the ``curse of dimensionality''. To overcome this problem, the sequential path planning (SPP) method, which assigns strict priorities to vehicles, was previously proposed; SPP allows multi-vehicle path planning to be done with a linearly-scaling computation complexity. In this work, we demonstrate the potential of SPP algorithm for large-scale systems. In particular, we simulate large-scale multi-vehicle systems in two different urban environments, a city environment and a multi-city environment, and use the SPP algorithm for trajectory planning. SPP is able to efficiently design collision-free trajectories in both environments despite the presence of disturbances in vehicles' dynamics. To ensure a safe transition of vehicles to their destinations, our method automatically allocates space-time reservations to vehicles while accounting for the magnitude of disturbances such as wind in a provably safe way. Our simulation results show an intuitive multi-lane structure in airspace, where the number of lanes and the distance between the lanes depend on the size of disturbances and other problem parameters.}

\begin{document}
\maketitle

% Introduction
% !TEX root = SPP_IoTjournal.tex
\section{Introduction \label{sec:introduction}}
Due to the recent surge of interest in the use of unmanned aerial systems (UASs) for civil applications such as package delivery, aerial surveillance, disaster response, among many others \cite{Tice91, Debusk10, Amazon16, AUVSI16, BBC16}, civilian airspace may in the near future contain up to thousands of unmanned aerial vehicles (UAVs), potentially in close proximity of humans, other UAVs, and other important assets. As a result, government agencies such as the Federal Aviation Administration (FAA) and National Aeronautics and Space Administration (NASA) of the United States are urgently trying to develop new scalable ways to organize an airspace in which potentially thousands of UAVs can fly together \cite{FAA13, Kopardekar16}.

One essential problem that needs to be addressed for this endeavor to be successful is that of trajectory planning: how a group of vehicles in the same vicinity can reach their destinations while avoiding situations which are considered dangerous, such as collisions. Many previous studies address this problem under different assumptions. In some studies, specific control strategies for the vehicles are assumed, and approaches such as those involving induced velocity obstacles \cite{Fiorini98, Chasparis05, Vandenberg08,Wu2012} and involving virtual potential fields to maintain collision avoidance \cite{Olfati-Saber2002, Chuang07} have been used. Methods have also been proposed for real-time trajectory generation \cite{Feng-LiLian2002}, for path planning for vehicles with linear dynamics in the presence of obstacles with known motion \cite{Ahmadzadeh2009}, and for cooperative path planning via waypoints which do not account for vehicle dynamics \cite{Bellingham}. Other related work include those which consider only the collision avoidance problem without path planning. These results include those that assume the system has a linear model \cite{Beard2003, Schouwenaars2004, Stipanovic2007}, rely on a linearization of the system model \cite{Massink2001, Althoff2011}, assume a simple positional state space \cite{Lin2015}, and many others \cite{Lalish2008, Hoffmann2008, Chen2016}.

However, to make sure that a dense group of UAVs can safely fly in the close vicinity of each other, we need the capability to flexibly plan provably safe and dynamically feasible trajectories without making strong assumptions on the vehicles' dynamics and other vehicles' motion. Moreover, any trajectory planning scheme that addresses collision avoidance must also guarantee both goal satisfaction and safety of UAVs despite disturbances caused by wind and communication faults \cite{Kopardekar16}. Finally, the proposed scheme should scale well with the number of vehicles, as well as result in an intuitive airspace structure for humans to monitor and potentially adjust.

The problem of trajectory planning and collision avoidance under disturbances in safety-critical systems has been well-studied using Hamilton-Jacobi (HJ) reachability analysis, which provides guarantees on goal satisfaction and safety of optimal system trajectories \cite{Barron90, Mitchell05, Bokanowski10, Bokanowski11, Margellos11, Fisac15}. Reachability-based methods are particularly suitable in the context of UAVs because of the hard guarantees that are provided. In reachability analysis, one computes the reach-avoid set, defined as the set of states from which the system can be driven to a target set while satisfying (possibly time-varying) state constraints at all times. A major practical appeal of this approach stems from the availability of modern numerical tools, which can compute various definitions of reachable sets \cite{Sethian96, Osher02, Mitchell02, Mitchell07b}. These numerical tools, for example, have been successfully used to solve a variety of differential games, path planning problems, and optimal control problems. Concrete practical applications include aircraft auto-landing \cite{Bayen07}, automated aerial refueling \cite{Ding08}, MPC control of quadrotors \cite{Bouffard12}, and multiplayer reach-avoid games \cite{Huang11}. Despite its power, the approach becomes numerically intractable as the state space dimension increases. In particular, reachable set computations involve solving a HJ partial differential equation (PDE) or variational inequality (VI) on a grid representing a discretization of the state space, resulting in an \textit{exponential} scaling of computational complexity with respect to the dimensionality of the problem. Therefore, dynamic programming-based approaches such as reachability analysis are not directly suitable for managing the next generation airspace, which is a large-scale system with a high-dimensional joint state space because of the high density of vehicles that needs to be accommodated \cite{Kopardekar16}.  

To overcome this problem, the Sequential Path Planning (SPP) method was proposed in \cite{Chen15c}. Here, vehicles are assigned a strict priority ordering. Higher-priority vehicles plan their paths without taking into account the lower-priority vehicles. Lower-priority vehicles treat higher-priority vehicles as moving obstacles. Under this assumption, time-varying formulations of reachability \cite{Bokanowski11, Fisac15} can be used to obtain the optimal and provably safe paths for each vehicle, starting from the highest-priority vehicle. Thus, the curse of dimensionality is overcome for the multi-vehicle path planning problem at the cost of a mild structural assumption, under which the computation complexity scales just \textit{linearly} with the number of vehicles. Intuitively, SPP algorithm allocates a space-time trajectory to each vehicle based on their priorities. The highest-priority vehicle gets to choose any space-time trajectory that does not collide with static obstacles, such as the optimal trajectory. The next vehicle's trajectory must not intersect with the trajectory of the highest-priority vehicle, and so on. Hence two vehicles can either follow same state tarjectory but at different times (referred to as \textit{time-separated} trajectories here on) or follow different state trajectories but at the same time (referred to as \textit{state-separated} trajectories here on), but not both. Finally, they can have different state trajectories at different times (referred to as \textit{state-time separated} trajectories here on). So by design, SPP algorithm ensures that the space-time trajectories of the vehicles do not intersect, and hence a safe transition to destination is guaranteed for all vehicles.  

Authors in \cite{Bansal2017} and \cite{chen2016robust}, respectively, extend SPP to the scenarios where disturbances and adversarial intruders are present in the system, resolving some of the practical challenges associated with the basic SPP algorithm in \cite{Chen15c}. The focus of these works, however, have mostly been on the theoretical development of SPP algorithm. Our focus in this work is instead on demonstrating the potential of SPP algorithm as a provably safe trajectory planning algorithm for large-scale systems. In particular, our main contributions in this paper are as follows:
\begin{itemize}
\item We simulate large-scale multi-vehicle systems in two different urban environments under the presence of disturbances in vehicles' dynamics. First, the SPP algorithm is used for trajectory planning at a city level and then at a regional level. For city level planning, we consider the city of San Francisco in California, USA, and for regional level planning we consider a part of San Francisco Bay Area in California, USA. The main differences between these two case studies are that the city level planning needs to take into account static physical obstacles such as tall buildings, whereas the origins and destinations are farther apart at the regional level. In both cases, we demonstrate that SPP algorithm is able to design provably-safe trajectories despite the disturbances.
\item We demonstrate how different types of space-time trajectories emerge naturally out of SPP algorithm between a given pair of origin and destination for different disturbance conditions and other problem parameters. These emerging behaviors, while being provably safe, are also intuitive and would facilitate human monitoring. 
\item We also show the reactivity of the control law obtained from SPP algorithm. In other words, we show that the obtained control law is able to effectively counteract the disturbances in real-time without requiring any communication with other vehicles.
\end{itemize}

The rest of the paper is organized as follows: in Section \ref{sec:formulation}, we formally present the SPP problem in the presence of disturbances. In Section \ref{sec:background}, we present a brief review of time-varying reachability, the basic SPP algorithm \cite{Chen15c} in the absence of disturbances, and the Robust Trajectory Tracking (RTT) method \cite{Bansal2017} to account for disturbances. We also use this algorithm for all our simulations in this paper. Simulation results are in Sections \ref{sec:city_sim} and \ref{sec:bayArea_sim}.

% Problem Formulation
% !TEX root = SPP_IoTjournal.tex
\section{Sequential Path Planning Problem \label{sec:formulation}}
Consider $\N$ vehicles (also denoted as \textit{SPP vehicles}) which participate in the SPP process $\veh_i, i = 1, \ldots, \N$. We assume their dynamics are given by

\begin{equation}
\label{eq:dyn}
\begin{aligned}
\dot\state_i &= \fdyn_i(\state_i, \ctrl_i, \dstb_i), t \le \sta_i \\
\ctrl_i &\in \cset_i, \dstb_i \in \dset_i, i = 1 \ldots, \N
\end{aligned}
\end{equation}

\noindent where $\state_i \in \R^{n_i}$, $\ctrl_i \in \cset_i$ and $\dstb_i \in \dset_i$, respectively, represent the state, control and disturbance experienced by vehicle $\veh_i$. We partition the state $\state_i$ into the position component $\pos_i \in \R^{n_\pos}$ and the non-position component $\npos_i \in \R^{n_i - n_\pos}$: $\state_i = (\pos_i, \npos_i)$. %We assume that the control functions $\ctrl_i(\cdot), \dstb_i(\cdot)$ are drawn from the set of measurable functions\footnote{A function $f:X\to Y$ between two measurable spaces $(X,\Sigma_X)$ and $(Y,\Sigma_Y)$ is said to be measurable if the preimage of a measurable set in $Y$ is a measurable set in $X$, that is: $\forall V\in\Sigma_Y, f^{-1}(V)\in\Sigma_X$, with $\Sigma_X,\Sigma_Y$ $\sigma$-algebras on $X$,$Y$.}. For convenience, 
We will use the sets $\cfset_i, \dfset_i$ to respectively denote the set of functions from which the control and disturbance functions $\ctrl_i(\cdot), \dstb_i(\cdot)$ are drawn.

% We further assume that the flow field $\fdyn_i: \R^{n_i}\times\cset_i\times\dset_i \rightarrow \R^{n_i}$ is uniformly continuous, bounded, and Lipschitz continuous in $\state_i$ for fixed $\ctrl_i$ and $\dstb_i$. With this assumption, given $\ctrl_i(\cdot) \in \cfset_i, \dstb_i(\cdot) \in \dfset_i$, there exists a unique trajectory solving \eqref{eq:dyn} \cite{EarlA.Coddington1955}. %We will denote trajectories of \eqref{eq:dyn} starting from state $\state^0_i$ at time $t_0$ under control $\ctrl_i(\cdot)$ and disturbance $\dstb_i(\cdot)$ as $\traj_i(t; \state^0_i, t_0, \ctrl_i(\cdot))$. Trajectories satisfy an initial condition and the differential equation \eqref{eq:dyn} almost everywhere:

%\begin{equation}
%\begin{aligned}
%\frac{d}{dt}\traj_i(t; \state^0_i, t_0, \ctrl_i(\cdot)) &= \fdyn(\state^0_i, \ctrl_i, \dstb_i) \\
%\traj_i(t_0; \state^0_i, t_0, \ctrl_i(\cdot)) &= \state^0_i
%\end{aligned}
%\end{equation}

%In addition, we assume that the disturbances $\dstb_i(\cdot)$ are drawn the set of non-anticipative strategies \cite{Mitchell05} $\Gamma$, defined as follows:
%\begin{equation}
%\begin{aligned}
%& \Gamma := \{\mathcal{N}: \cfset_i \rightarrow \dfset_i:  \ctrl_i(r) = \hat{\ctrl}_i(r) \text{ a. e. } r\in[t,s] \\
%& \Rightarrow \mathcal{N}[\ctrl_i](r) = \mathcal{N}[\hat{\ctrl}_i](r) \text{ a. e. } r\in[t,s]\}
%\end{aligned}
%\end{equation}

Each vehicle $\veh_i$ has initial state $\state^0_i$, and aims to reach its target $\targetset_i$ by some scheduled time of arrival $\sta_i$. The target in general represents some set of desirable states, for example the destination of $\veh_i$. %For most of the paper we will make the assumption that $\edt_i$ is early enough for $\veh_i$ to feasibly get to $\targetset_i$ on time; this can be done by letting $\edt_i \rightarrow -\infty$. The assumption on $\edt_i$ is merely for convenience: in situations where $\edt_i$ is $-\infty$. In some situations, we may find that it is infeasible for $\veh_i$ to get to $\targetset_i$ at or before $\sta_i$. Whenever unsure, we may first determine the earliest feasible $\sta_i$ as described in Section \ref{sec:intruder}.
On its way to $\targetset_i$, $\veh_i$ must avoid a set of static obstacles $\soset_i \subset \R^{n_i}$. The interpretation of $\soset_i$ could be a tall building or any set of states that are forbidden for each SPP vehicle. In addition to the static obstacles, each vehicle $\veh_i$ must also avoid the danger zones with respect to every other vehicle $\veh_j, j\neq i$. The danger zones in general can represent any joint configurations between $\veh_i$ and $\veh_j$ that are considered to be unsafe. We define the danger zone of $\veh_i$ with respect to $\veh_j$ to be
\begin{equation}
\dz_{ij} = \{(\state_i, \state_j): \|\pos_i - \pos_j\|_2 \le \rc\}
\end{equation}
\noindent whose interpretation is that $\veh_i$ and $\veh_j$ are considered to be in an unsafe configuration when they are within a distance of $\rc$ of each other. In particular, $\veh_i$ and $\veh_j$ are said to have collided, ifif $(\state_i, \state_j) \in \dz_{ij}$.

Given the set of SPP vehicles, their targets $\targetset_i$, the static obstacles $\soset_i$, and the vehicles' danger zones with respect to each other $\dz_{ij}$, our goal is, for each vehicle $\veh_i$, to synthesize a controller which guarantees that $\veh_i$ reaches its target $\targetset_i$ at or before the scheduled time of arrival $\sta_i$, while avoiding the static obstacles $\soset_i$ as well as the danger zones with respect to all other vehicles $\dz_{ij}, j\neq i$. In addition, we would like to obtain the latest departure time $\ldt_i$ such that $\veh_i$ can still arrive at $\targetset_i$ on time.

In general, the above optimal path planning problem must be solved in the joint space of all $\N$ SPP vehicles. However, due to the high joint dimensionality, a direct dynamic programming-based solution is intractable. Therefore, authors in \cite{Chen15c} proposed to assign a priority to each vehicle, and perform SPP given the assigned priorities. Without loss of generality, let $\veh_j$ have a higher priority than $\veh_i$ if $j<i$. Under the SPP scheme, higher-priority vehicles can ignore the presence of lower-priority vehicles, and perform path planning without taking into account the lower-priority vehicles' danger zones. A lower-priority vehicle $\veh_i$, on the other hand, must ensure that it does not enter the danger zones of the higher-priority vehicles $\veh_j, j<i$; each higher-priority vehicle $\veh_j$ induces a set of time-varying obstacles $\ioset_i^j(t)$, which represents the possible states of $\veh_i$ such that a collision between $\veh_i$ and $\veh_j$ could occur.

It is straight-forward to see that if each vehicle $\veh_i$ is able to plan a trajectory that takes it to $\targetset_i$ while avoiding the static obstacles $\soset_i$ and the danger zones of \textit{higher-priority vehicles} $\veh_j, j<i$, then the set of SPP vehicles $\veh_i, i=1,\ldots,\N$ would all be able to reach their targets safely. %With the SPP scheme, the additional structure provided by the vehicle priorities allows us to reduce the complexity of the joint path planning problem. As we will see, 
Under the SPP scheme, path planning can be done sequentially in descending order of vehicle priority in the state space of only a single vehicle. Thus, SPP provides a solution whose complexity scales linearly with the number of vehicles, as opposed to exponentially with a direct application of dynamic programming approaches. %In the presence of a single intruder, the computation complexity scaling becomes quadratic.

% Background material
% !TEX root = ./SPP_IoTjournal.tex
\section{Background \label{sec:background}}
In this section, we present the basic SPP algorithm \cite{Chen15c} in which disturbances are ignored and perfect information of vehicles’ positions is assumed. This simplification allows us to clearly present the basic SPP algorithm. However, in presence of disturbances, it is no longer possible to commit to exact trajectories (and hence positions), since the disturbance $\dstb_i(\cdot)$ is \textit{a priori} unknown. Thus, disturbances and incomplete information significantly complicate the SPP scheme. We next present the robust trajectory tracking algorithm \cite{Bansal2017} that can be used to make basic SPP approach robust to disturbances as well as to an imperfect knowledge of other vehicles' positions. All of these algorithms use time-varying reachability analysis to provide goal satisfaction and safety guarantees; therefore, we start with an overview of time-varying reachability.
%
%
%The basic SPP algorithm presented in \cite{Chen15c} ignores disturbances in vehicles' dynamics and assumes perfect information of vehicles' positions. However, in presence of disturbances, it is no longer possible to commit to exact trajectories (and hence positions), since the disturbance $\dstb_i(\cdot)$ is \textit{a priori} unknown. Thus, disturbances and incomplete information significantly complicate the SPP scheme. In this section, we present the robust trajectory tracking algorithm (proposed in \cite{Bansal2017}) that can be used to make basic SPP approach robust to disturbances as well as to an imperfect knowledge of other vehicles' positions. In the next section, we will present an algorithm to further robustify the SPP approach by considering how the set of SPP vehicles may respond to the presence of an intruder vehicle which may be adversarial. All of these algorithms use time-varying reachability analysis to provide goal satisfaction and safety guarantees; therefore, we start with an overview of time-varying reachability.
%
%In this section, we present the SPP algorithm under different assumptions. We begin with the basic SPP algorithm in which disturbances are ignored and perfect information of vehicles' positions is assumed. This simplification allows us to clearly present the basic SPP algorithm. Next, we present the robust trajectory tracking algorithm that can be used to make basic SPP approach robust to disturbances as well as an imperfect knowledge of other vehicles' positions. 

% !TEX root = ../SPP_IoTjournal.tex
\subsection{Time-Varying Reachability Background \label{sec:HJIVI}}
We will be using reachability analysis to compute a backward reachable set (BRS) $\brs$ given some target set $\targetset$, time-varying obstacle $\obsset(t)$, and the Hamiltonian function $\ham$ which captures the system dynamics as well as the roles of the control and disturbance. The BRS $\brs$ in a time interval $[t, t_f]$ will be denoted by

\begin{equation}
\brs(t, t_f) \quad \text{ (backward reachable set)}
\end{equation}

%In the SPP scheme, a lower-priority vehicle must avoid a set of moving obstacles on its way to the target. Several formulations of reachability are able to perform optimal path planning with hard guarantees on safety and performance under disturbances in such a scenario \cite{Bokanowski11, Fisac15}. 
Several formulations of reachability are able to account for time-varying obstacles \cite{Bokanowski11, Fisac15} (or state constraints in general). For our application in SPP, we utilize the time-varying formulation in \cite{Fisac15}, which accounts for the time-varying nature of systems without requiring augmentation of the state space with the time variable. In the formulation in \cite{Fisac15}, a BRS is computed by solving the following \textit{final value} double-obstacle HJ VI:

\begin{equation}
\label{eq:HJIVI_BRS}
\begin{aligned}
\max \Big\{ \min \{&D_t \valfunc(t, \state) + \ham(t, \state, \nabla \valfunc(t, \state)), \fc(\state) - \valfunc(t, \state) \}, \\
&-\obsfunc(t, \state) - \valfunc(t, \state) \Big\} = 0, \quad t \le t_f \\
&\valfunc(t_f, \state) = \max\{\fc(\state), -\obsfunc(t_f, \state)\}
\end{aligned}
\end{equation}

%In a similar fashion, the FRS is computed by solving the following \textit{initial value} HJ PDE:
%
%\begin{equation}
%\label{eq:HJIVI_FRS}
%\begin{aligned}
%D_t \valfuncfwd(t, \state) + &\ham(t, \state, \nabla \valfuncfwd(t, \state)) = 0 , \quad t \ge t_0  \\
%&\valfuncfwd(t_0, \state) = \max\{\fc(\state), -\obsfunc(t_0, \state)\}
%\end{aligned}
%\end{equation}
%
In \eqref{eq:HJIVI_BRS}, the function $\ic(\state)$ is the implicit surface function representing the target set $\targetset = \{\state: \ic(\state) \le 0\}$. Similarly, the function $\obsfunc(t, \state)$ is the implicit surface function representing the time-varying obstacles $\obsset(t) = \{\state: \obsfunc(t,\state)\le 0\}$. The BRS $\brs(t, t_f)$ is given by

%\begin{equation}
%\label{eq:implicitValfuncs}
%\begin{aligned}
%\brs(t, t_f) &= \{\state: \valfunc(t, \state) \le 0\} \\
%\frs(t_0, t) &= \{\state: \valfuncfwd(t, \state) \le 0 \}
%\end{aligned}
%\end{equation}
\begin{equation}
\label{eq:implicitValfuncs}
\brs(t, t_f) = \{\state: \valfunc(t, \state) \le 0\}
\end{equation}

Some of the reachability computations will not involve an obstacle set $\obsset(t)$, in which case we can simply set $\obsfunc(t, \state) \equiv \infty$ which effectively means that the outside maximum is ignored in \eqref{eq:HJIVI_BRS}. %Also, note that unlike in \eqref{eq:HJIVI_BRS}, there is no inner minimization in \eqref{eq:HJIVI_FRS}. As we will see later, we will be using the BRS to determine all states that can reach some target set \textit{within the time horizon} $[t,t_f]$, whereas we will be using the FRS to determine where a vehicle could be \textit{at some particular time} $t$. In addition, \eqref{eq:HJIVI_FRS} has no outer maximum, since the FRSs that we will compute will not involve any obstacles.

%\MCnote{Not sure if this is needed, but putting it here for now} For clarity, sometimes we will denote value functions computed using the target set $\targetset$, obstacle $\obsset(t)$, and Hamiltonian $\ham$ as 
%
%\begin{equation}
%\begin{aligned}
%\valfunc(t, \state; \targetset, \obsset(\cdot), \ham) &\quad \text{(Value function representing BRS)} \\
%\valfuncfwd(t, \state; \targetset, \obsset(\cdot), \ham) &\quad \text{(Value function representing FRS)}
%\end{aligned}
%\end{equation}

The Hamiltonian, $\ham(t, \state, \nabla \valfunc(t,\state))$, depends on the system dynamics, and the role of control and disturbance. Whenever $\ham$ does not depend explicit on $t$, we will drop the argument. In addition, the Hamiltonian is an optimization that produces the optimal control $\ctrl^*(t, \state)$ and optimal disturbance $\dstb^*(t, \state)$, once $\valfunc$ is determined. For BRSs, whenever the existence of a control (``$\exists \ctrl$'') or disturbance is sought, the optimization is a minimum over the set of controls or disturbance. Whenever a BRS characterizes the behavior of the system for all controls (``$\forall \ctrl$'') or disturbances, the optimization is a maximum. We will introduce precise definitions of reachable sets, expressions for the Hamiltonian, expressions for the optimal controls as needed for the many different reachability calculations we use. %As an example, suppose we are given a generic dynamical system $\dot \state = \fdyn(t, \state, \ctrl, \dstb)$ in which the control $\ctrl(\cdot)$ aims to reach the target $\targetset$ while the disturbance aims to keep the system away from the target, then
%
%\begin{equation}
%\label{eq:ham_ex}
%\begin{aligned}
%\ham(\state, \nabla \valfunc(t,\state)) = \min_\ctrl \max_\dstb \nabla \valfunc(t,\state) \cdot \fdyn(t, \state, \ctrl, \dstb)
%\end{aligned}
%\end{equation}
%
%In addition, given the Hamiltonian \eqref{eq:ham_ex}, the optimal control and disturbance can be obtained as the optimum of the optimization:
%
%\begin{equation}
%\label{eq:opt_ctrl_dstb_ex}
%\begin{aligned}
%\ctrl^*(t) &= \arg \max \min_\dstb \nabla \valfunc(t,\state) \cdot \fdyn(\state, \ctrl, \dstb)\\
%\dstb^*(t) &= \arg \min \nabla \valfunc(t,\state) \cdot \fdyn(\state, \ctrl^*(t), \dstb)
%\end{aligned}
%\end{equation}
%
%In anticipation of the many different reachable sets that will be computed, we will in general denote the optimal control and disturbance derived from a BRS $\brs(t; \targetset, \obsset(\cdot), \ham)$ given the target set $\targetset(t)$, obstacles $\obsset(t)$, and Hamiltonian $\ham$ as 
%
%\begin{equation}
%\label{eq:opt_ctrl}
%\begin{aligned}
%\ctrl^*(t, \state_i; \targetset, \obsset(\cdot), \ham) &\quad \text{(optimal control)} \\
%\dstb^*(t, \state_i; \targetset, \obsset(\cdot), \ham) &\quad \text{(optimal disturbance)}
%\end{aligned}
%\end{equation}
%
%\textcolor{red}{Only keep time and/or state arguments when notation gets too heavy; use subscripts to specify the different sets instead.}

% !TEX root = ../SPP_IoTjournal.tex
\subsection{SPP Without Disturbances and Intruder\label{sec:basic}}
In this section, we give an overview of the basic SPP algorithm assuming that there is no disturbance and intruder affecting the vehicles. Although in practice, such assumptions do not hold, the description of the basic SPP algorithm will introduce the notation needed for describing the subsequent, more realistic versions of SPP. The majority of the content in this section is taken from \cite{Chen15c}.

Recall that the SPP vehicles $\veh_i, i=1,\ldots,N$, are each assigned a strict priority, with $\veh_j$ having a higher priority than $\veh_i$ if $j<i$. In the absence of disturbances, we can write the dynamics of the SPP vehicles as
\begin{equation}
\label{eq:dyn_no_dstb}
\begin{aligned}
\dot\state_i &= \fdyn_i(\state_i, \ctrl_i), t \le \sta_i \\
\ctrl_i &\in \cset_i, \qquad i = 1 \ldots, \N
\end{aligned}
\end{equation}

In SPP, each vehicle $\veh_i$ plans the path to its target set $\targetset_i$ while avoiding static obstacles $\soset_i$ and the obstacles $\ioset_i^j(t)$ induced by higher-priority vehicles $\veh_j, j<i$. Path planning is done sequentially starting from the first vehicle and proceeding in descending priority, $\veh_1, \veh_2, \ldots, \veh_{\N}$ so that each of the path planning problems can be done in the state space of only one vehicle. During its path planning process, $\veh_i$ ignores the presence of lower-priority vehicles $\veh_k, k>i$, and induces the obstacles $\ioset_k^i(t)$ for $\veh_k, k>i$.

From the perspective of $\veh_i$, each of the higher-priority vehicles $\veh_j, j<i$ induces a time-varying obstacle denoted $\ioset_i^j(t)$ that $\veh_i$ needs to avoid. Therefore, each vehicle $\veh_i$ must plan its path to $\targetset_i$ while avoiding the union of all the induced obstacles as well as the static obstacles. Let $\obsset_i(t)$ be the union of all the obstacles that $\veh_i$ must avoid on its way to $\targetset_i$:
\begin{equation}
\label{eq:obsseti}
\obsset_i(t)  = \soset_i \cup \bigcup_{j=1}^{i-1} \ioset_i^j(t)
\end{equation}

With full position information of higher priority vehicles, the obstacle induced for $\veh_i$ by $\veh_j$ is simply
\begin{equation}
\label{eq:ioset_basic}
\ioset_i^j(t) = \{\state_i: \|\pos_i - \pos_j(t)\|_2 \le \rc \}
\end{equation}

Each higher priority vehicle $\veh_j$ plans its path while ignoring $\veh_i$. Since path planning is done sequentially in descending order or priority, the vehicles $\veh_j, j<i$ would have planned their paths before $\veh_i$ does. Thus, in the absence of disturbances, $\pos_j(t)$ is \textit{a priori} known, and therefore $\ioset_i^j(t), j<i$ are known, deterministic moving obstacles, which means that $\obsset_i(t)$ is also known and deterministic. Therefore, the path planning problem for $\veh_i$ can be solved by first computing the BRS $\brs_i^\text{basic}(t, \sta_i)$, defined as follows:
\begin{equation}
\label{eq:BRS_basic}
\begin{aligned}
\brs_i^\text{basic}(t, \sta_i) = & \{y: \exists \ctrl_i(\cdot) \in \cfset_i, \state_i(\cdot) \text{ satisfies \eqref{eq:dyn_no_dstb}}, \\
& \forall s \in [t, \sta_i],\state_i(s) \notin \obsset_i(s), \\
& \exists s \in [t, \sta_i], \state_i(s) \in \targetset_i, \state_i(t) = y\}
\end{aligned}
\end{equation}

The BRS $\brs(t, \sta_i)$ can be obtained by solving \eqref{eq:HJIVI_BRS} with $\targetset = \targetset_i$, $\obsset(t) = \obsset_i(t)$, and the Hamiltonian 
\begin{equation}
\label{eq:basicham}
\ham_i^\text{basic}(\state_i, \costate) = \min_{\ctrl_i\in\cset_i} \costate \cdot \fdyn_i(\state_i, \ctrl_i)
\end{equation}

The optimal control for reaching $\targetset_i$ while avoiding $\obsset_i(t)$ is then given by
\begin{equation}
\label{eq:basicOptCtrl}
\ctrl_i^\text{basic}(t, \state_i) = \arg \min_{\ctrl_i\in\cset_i} \costate \cdot \fdyn_i(\state_i, \ctrl_i)
\end{equation}
\noindent from which the trajectory $\state_i(\cdot)$ can be computed by integrating the system dynamics, which in this case are given by \eqref{eq:dyn_no_dstb}. In addition, the latest departure time $\ldt_i$ can be obtained from the BRS $\brs(t, \sta_i)$ as $\ldt_i = \arg \sup_t \{\state_i^0 \in \brs(t, \sta_i)\}$. In summary, the basic SPP algorithm is given as follows:
\begin{alg}
\label{alg:basic}
\textbf{Basic SPP algorithm}: Suppose we are given initial conditions $\state_i^0$, vehicle dynamics \eqref{eq:dyn_no_dstb}, target sets $\targetset_i$, and static obstacles $\soset_i, i = 1\ldots, \N$. For each $i$ in ascending order starting from $i=1$ (which corresponds to descending order of priority),
\begin{enumerate}
\item determine the total obstacle set $\obsset_i(t)$, given in \eqref{eq:obsseti}. In the case $i=1$, $\obsset_i(t) = \soset_i ~ \forall t$;
\item compute the BRS $\brs_i^\text{basic}(t, \sta_i)$ defined in \eqref{eq:BRS_basic}. The latest departure time $\ldt_i$ is then given by $\arg \sup_t \{\state^0_i \in \brs_i^\text{basic}(t, \sta_i)\}$;
\item determine the trajectory $\state_i(\cdot)$ using vehicle dynamics \eqref{eq:dyn_no_dstb}, with the optimal control  $\ctrl_i^\text{basic}(\cdot)$ given by \eqref{eq:basicOptCtrl};
\item given $\state_i(\cdot)$, compute the induced obstacles $\ioset_k^i(t)$ for each $k>i$. In the absence of disturbances, $\ioset_k^i(t)$ is given by \eqref{eq:ioset_basic}.
\end{enumerate}
\end{alg}
% !TEX root = ../SPP_IoTjournal.tex
\subsection{Robust Trajectory Tracking (RTT) \label{sec:rtt}}
In the basic SPP algorithm, lower priority vehicles know the trajectories of all higher priority vehicles. The region that a lower priority vehicle needs to avoid is thus simply given by the danger zones around these trajectories; however, disturbances and incomplete information significantly complicate the SPP scheme. Committing to exact trajectories is no longer possible, since the disturbance $\dstb_i(\cdot)$ is \textit{a priori} unknown. Thus, the induced obstacles $\ioset_i^j(t)$ are no longer just the danger zones centered around positions. In this section, we provide an overview of the RTT algorithm that can overcome these issues. For simplicity of explanation, we will assume that no static obstacles exist, but method can be generalized even when static obstacles do exist. The material in this section is taken partially from \cite{Bansal2017}. Note that other algorithms have been developed in \cite{Bansal2017} to account for the disturbances, we use RTT algorithm for the simulations in this paper and only present RTT algorithm here. Interested readers are referred to \cite{Bansal2017} for the other algorithms. 

Even though it is impossible to commit to and track an exact trajectory in the presence of disturbances, it may still be possible to instead \textit{robustly} track a feasible \textit{nominal} trajectory with a bounded error at all times. If this can be done, then the tracking error bound can be used to determine the induced obstacles. Here, computation is done in two phases: the \textit{planning phase} and the \textit{disturbance rejection phase}. 

In the planning phase, a nominal trajectory $\state_{r,j}(\cdot)$ is computed that is feasible in the absence of disturbances. This planning is done for a reduced control set $\cset^p\subset\cset$, as some margin is needed to reject unexpected disturbances while tracking the nominal trajectory.

%In the disturbance rejection phase, we compute a bound on the tracking error.%\MCnote{don't need to explain where error comes bound, imo}%, caused by a vehicle's inability to exactly track the nominal trajectory in the presence of disturbances. 

%It is important to note that the planning phase does not make full use of a vehicle's control authority, as some margin is needed to reject unexpected disturbances while tracking the nominal trajectory. Therefore, in this method, planning is done for a reduced control set $\cset^p\subset\cset$. The resulting trajectory reference will not utilize the vehicle's full control capability; additional maneuverability is available at execution time to counteract external disturbances.

In the disturbance rejection phase, we compute a bound on the tracking error, independently of the nominal trajectory. To compute this error bound, we find a robust controlled-invariant set in the joint state space of the vehicle and a tracking reference that may ``maneuver" arbitrarily in the presence of an unknown bounded disturbance. Taking a worst-case approach, the tracking reference can be viewed as a virtual evader vehicle that is optimally avoiding the actual vehicle to enlarge the tracking error. We therefore can model trajectory tracking as a pursuit-evasion game in which the actual vehicle is playing against the coordinated worst-case action of the virtual vehicle and the disturbance. %In general, this game will be governed by dynamics of the form:

%\begin{equation}
%\label{eq:jdyn} % Joint dynamics
%\begin{aligned}
%\dot{\state_j} &= f_j(t, \state_j, \ctrl_j, \dstb_j), \quad \dot{x_r} =f_j(t,x_r,\ctrl_r,0),\\
%\ctrl_j &\in \cset_j, \ctrl_r\in\cset^\pos_j, d \in \dset_j, \quad t \in [0, T]
%\end{aligned}
%\end{equation}
%where 

Let $\state_j$ and $\state_{r,j}$ denote the states of the actual vehicle $\veh_j$ and the virtual evader, respectively, and define the tracking error $e_j=\state_j-\state_{r,j}$. When the error dynamics are independent of the absolute state as in \eqref{eq:edyn} (and also (7) in \cite{Mitchell05}), we can obtain error dynamics of the form

\begin{equation}
\label{eq:edyn} % Error dynamics
\begin{aligned}
\dot{e_j} &= \fdyn_{e_j}(e_j, \ctrl_j, \ctrl_{r,j},\dstb_j), \\
\ctrl_j &\in \cset_j, \ctrl_{r,j} \in \cset^p_j, \dstb_j \in \dset_j, \quad t \leq 0
\end{aligned}
\end{equation}

To obtain bounds on the tracking error, we first conservatively estimate the error bound around any reference state $\state_{r,j}$, denoted $\errorbound_j$:

\begin{equation} \label{eqn:err}
\errorbound_j = \{e_j: \|\pos_{e_j}\|_2 \le R_{\text{EB}} \}, 
\end{equation}

\noindent where $\pos_{e_j}$ denotes the position coordinates of $e_j$ and $R_{\text{EB}}$ is a design parameter. We next solve a reachability problem with its complement $\errorbound_j^c$, the set of tracking errors violating the error bound, as the target in the space of the error dynamics. From $\errorbound_j^c$, we compute the following BRS:

\begin{equation} \label{eqn:errBound}
\begin{aligned}
\brs^{\text{EB}}_{j}(t, 0) = & \{y: \forall \ctrl_j(\cdot) \in \cfset_j, \exists \ctrl_{r, j}(\cdot) \in \cfset^\pos_j, \exists \dstb_j(\cdot) \in \dfset_i, \\
& e_j(\cdot) \text{ satisfies \eqref{eq:edyn}}, e_j(t) = y, \\
& \exists s \in [t, 0], e_j(s) \in \errorbound_j^c\}, 
\end{aligned}
\end{equation}
where the Hamiltonian to compute the BRS is given by:
\begin{equation}
\begin{aligned}
H^{\text{EB}}_{j}(e_j, \costate) &= \max_{\ctrl_j \in \cset_j} \min_{\ctrl_r \in \cset^\pos_j, \dstb_j \in \dset_j} \costate \cdot \fdyn_{e_j}(e_j, \ctrl_j, \ctrl_{r,j}, \dstb_j).
\end{aligned}
\end{equation}

Letting $t \to -\infty$, we obtain the infinite-horizon control-invariant set $\disckernel_j := \lim_{t \to -\infty} \left( \brs^{\text{EB}}_{j}(t, 0) \right)^c$. If $\disckernel_j$ is nonempty, then the tracking error $e_j$ at flight time is guaranteed to remain within $\disckernel_j \subseteq \errorbound_j$ provided that the vehicle starts inside $\disckernel_j$ and subsequently applies the feedback control law

\begin{equation}
\label{eq:robust_tracking_law}
\tracklaw_j(e_j) = \arg\max_{\ctrl_j \in \cset_j} \min_{\ctrl_r \in\cset^\pos_j, \dstb_j \in \dset_j} \costate \cdot \fdyn_{e_j}(e_j,\ctrl_j,\ctrl_{r, j},\dstb_j).
\end{equation}

The induced obstacles by each higher-priority vehicle $\veh_j$ can thus be obtained by: 
\begin{equation} 
\label{eqn:rttObs}
\begin{aligned}
\ioset_i^j(t) &=  \{\state_i: \exists y \in \pfrs_j(t), \|\pos_i - y\|_2 \le \rc \} \\
\pfrs_j(t) &= \{\pos_j: \exists \npos_j, (\pos_j, \npos_j) \in \boset_j(t)\} \\
\boset_j(t) &= \disckernel_j  + \state_{r,j}(t),
\end{aligned}
\end{equation}

\noindent where the ``$+$'' in \eqref{eqn:rttObs} denotes the Minkowski sum\footnote{The Minkowski sum of sets $A$ and $B$ is the set of all points that are the sum of any point in $A$ and $B$.}. Intuitively, if $\veh_j$ is tracking $\state_{r,j}(t)$, then it will remain within the error bound $\disckernel_j$ around $\state_{r,j}(t) ~\forall t$. This is precisely the set $\pfrs_j(t)$. The induced obstacles can then be obtained by augmenting a danger zone around this set. Finally, we can obtain the total obstacle set $\obsset_i(t)$ using:
\begin{equation}
\label{eq:obsseti}
\obsset_i(t)  = \soset_i \cup \bigcup_{j=1}^{i-1} \ioset_i^j(t)
\end{equation} 

Since each vehicle $\veh_j$, $j<i$, can only be guaranteed to stay within $\disckernel_j$, we must make sure during the path planning of $\veh_i$ that at any given time, the error bounds of $\veh_i$ and $\veh_j$, $\disckernel_i$ and $\disckernel_j$, do not intersect. This can be done by augmenting the total obstacle set by $\disckernel_i$:%This can be done by choosing the induced obstacle to be the Minkowski sum\footnote{The Minkowski sum of sets $A$ and $B$ is the set of all points that are the sum of any point in $A$ and $B$.} of the error bounds. Thus,

\begin{equation} 
\label{eqn:rttAugObs}
\tilde{\obsset}_i(t) = \obsset_i(t) + \disckernel_i.
\end{equation}

Finally, given $\disckernel_i$, we can guarantee that $\veh_i$ will reach its target $\targetset_i$ if $\disckernel_i \subseteq \targetset_i$; thus, in the path planning phase, we modify $\targetset_i$ to be $\tilde{\targetset}_i := \{\state_i: \disckernel_i + \state_i \subseteq \targetset_i\}$, and compute a BRS, with the control authority $\cset^\pos_i$, that contains the initial state of the vehicle. Mathematically,

\begin{equation}
\label{eq:rttBRS}
\begin{aligned}
\brs_i^\text{rtt}(t, \sta_i) = & \{y: \exists \ctrl_i(\cdot) \in \cfset^p_i, \state_i(\cdot) \text{ satisfies \eqref{eq:dyn_no_dstb}},\\
&\forall s \in [t, \sta_i], \state_i(s) \notin \tilde{\obsset}_i(t), \\
& \exists s \in [t, \sta_i], \state_i(s) \in \tilde{\targetset}_i, \state_i(t) = y\}
\end{aligned}
\end{equation}

The BRS $\brs_i^\text{rtt}(t, \sta_i)$ can be obtained by solving \eqref{eq:HJIVI_BRS} using the Hamiltonian: 
\begin{equation}
\label{eq:RTTham}
\ham_i^\text{rtt}(\state_i, \costate) = \min_{\ctrl_i \in \cset^\pos_i } \costate \cdot \fdyn_i(\state_i, \ctrl_i)
\end{equation}

The corresponding optimal control for reaching $\tilde{\targetset}_i$ is given by:
\begin{equation}
\label{eq:RTTOptCtrl}
\ctrl_i^\text{rtt}(t) = \arg \min_{\ctrl_i \in \cset^\pos_i } \costate \cdot \fdyn_i(\state_i, \ctrl_i).
\end{equation}

The nominal trajectory $\state_{r,i}(\cdot)$ can thus be obtained by using vehicle dynamics in the absence of disturbances
\begin{equation}
\label{eq:dyn_no_dstb}
\begin{aligned}
\dot\state_i &= \fdyn_i(\state_i, \ctrl_i), t \le \sta_i \\
\ctrl_i &\in \cset_i, \qquad i = 1 \ldots, \N,
\end{aligned}
\end{equation}

with the optimal control  $\ctrl_i^\text{rtt}(\cdot)$ given by \eqref{eq:RTTOptCtrl}. From the resulting nominal trajectory $\state_{r,i}(\cdot)$, the overall control policy to reach $\targetset_i$ can be obtained via \eqref{eq:robust_tracking_law}. The robust trajectory tracking method can be summarized as follows:
\begin{alg}
\label{alg:rtt}
\textbf{Robust trajectory tracking algorithm}: Given initial conditions $\state_i^0$, vehicle dynamics \eqref{eq:dyn}, target sets $\targetset_i$, and static obstacles $\soset_i, i = 1\ldots, \N$, for each $i$,
\begin{enumerate}
\item determine the total obstacle set $\obsset_i(t)$, given in \eqref{eq:obsseti}. In the case $i=1$, $\obsset_i(t) = \soset_i ~ \forall t$;
\item decide on a reduced control authority $\cset^\pos_i$ for the planning phase, and choose a parameter $R_{\text{EB}}$ to conservatively bound the tracking error;
\item compute the BRS $\brs^{\text{EB}}_{i}(t, 0)$ using \eqref{eqn:errBound} and make sure that $\disckernel_i \neq \emptyset$. Given $R_{\text{EB}}$, the error bound on the tracking error is given by $\disckernel_i$;
\item using $\disckernel_i$, determine the augmented obstacle set $\tilde{\obsset}_i(t)$, given in \eqref{eqn:rttAugObs};
\item compute the BRS $\brs_i^\text{rtt}(t, \sta_i)$ as described in \eqref{eq:rttBRS} using the reduced target set $\tilde{\targetset}_i$, $\tilde{\obsset}_i(t)$ as obstacles, and the control authority $\cset^\pos_i$. The latest departure time $\ldt_i$ is then given by $\arg \sup_t \state^0_i \in \brs_i^\text{rtt}(t, \sta_i)$;
\item compute the nominal trajectory $\state_{r,i}(\cdot)$ for $\veh_i$ in the absence of disturbances, which can be obtained using the vehicle dynamics in \eqref{eq:dyn_no_dstb} and the optimal control given in \eqref{eq:RTTOptCtrl};
\item the induced obstacles $\ioset_k^i(t)$ for each $k>i$ can be computed using $\disckernel_i$ and $\state_{r,i}(\cdot)$ via \eqref{eqn:rttObs}.
\end{enumerate}
\end{alg}

\section{City Environment Simulation \label{sec:city_sim}}
We now illustrate SPP algorithm using a 50-vehicle UAV system where UAVs are flying through a city environment. This setup can be representative of many UAV applications, such as package delivery, aerial surveillance, etc. 
% !TEX root = ../../SPP_IoTjournal.tex
\subsection{Setup \label{sec:city_simSetup}}
\begin{figure}[H]
  \centering
  \includegraphics[width=\columnwidth]{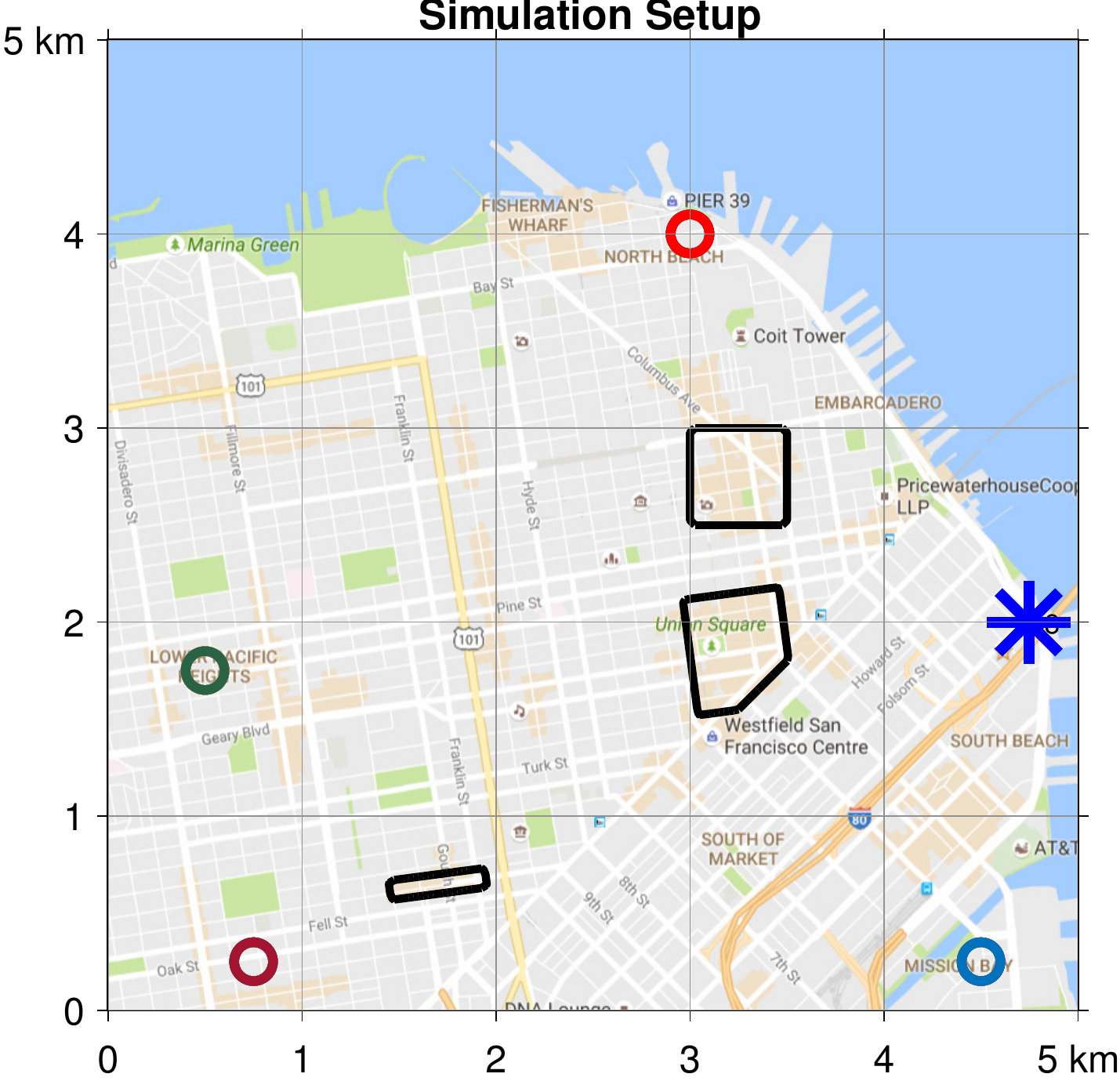}
  \caption{Simulation setup. A $25$ km$^2$ area in the City of San Francisco is used as the space for the SPP vehicles. SPP vehicles originate from the blue star and go to one of the four destinations, denoted by circles. Tall buildings in the downtown area are used as static obstacles, represented by the black contours.}
  \label{fig:sf_setup}
\end{figure}

We grid the City of San Francisco (SF) in California, USA, and use it as our position space, as shown in Figure \ref{fig:sf_setup}. 
Each box in Figure \ref{fig:sf_setup} represents a $1$ km$^2$ area of SF. The origin point for the vehicles is denoted by the blue star. Four different areas in the city are chosen as the destinations for the vehicles. Mathematically, the target sets $\targetset_i$ of the vehicles are circles of radius $r$ in the position space, i.e. each vehicle is trying to reach some desired set of positions. In terms of the state space $\state_i$, the target sets are defined as
\begin{equation}
\label{eq:target_sim}
\targetset_i = \{\state_i: \|\pos_i - c_i\|_2 \le r\}
\end{equation}
\noindent where $c_i$ are centers of the target circles. In this simulation, we use $r = 100$ m. The four targets are represented by four circles in Figure \ref{fig:sf_setup}. The destination of each vehicle is chosen randomly from these four destinations. Finally, some areas in downtown SF and the city hall are used as representative static obstacles for the SPP vehicles, denoted by black contours in Figure \ref{fig:sf_setup}.

For this simulation, we use the following dynamics for each vehicle:
\begin{equation}
\label{eq:dyn_i}
\begin{aligned}
\dot{\pos}_{x,i} &= v_i \cos \theta_i + \dstb_{x,i} \\
\dot{\pos}_{y,i} &= v_i \sin \theta_i + \dstb_{y,i}\\
\dot{\theta}_i &= \omega_i, \\
& \underline{v} \le v_i \le \bar{v}, ~|\omega_i| \le \bar{\omega}, \\
& \|(\dstb_{x,i}, \dstb_{y,i}) \|_2 \le d_{r}
\end{aligned}
\end{equation}
\noindent where $\state_i = (\pos_{x,i}, \pos_{y,i}, \theta_i)$ is the state of vehicle $\veh_i$, $\pos_i = (\pos_{x,i}, \pos_{y,i})$ is the position, $\theta_i$ is the heading, and $d = (d_{x,i}, d_{y,i})$ represents $\veh_i$'s disturbances, for example wind, that affect its position evolution. The control of $\veh_i$ is $u_i = (v_i, \omega_i)$, where $v_i$ is the speed of $\veh_i$ and $\omega_i$ is the turn rate; both controls have a lower and upper bound. To make our simulations as close as possible to real scenarios, we choose velocity and turn-rate bounds as $\underline{v} = 0$ m/s, $\bar{v} = 25$ m/s, $\bar\omega = 2$ rad/s, aligned with the modern UAV specifications \cite{UAVspecs1, UAVspecs2}. The disturbance bounds are chosen to be either $\dstb_r = 6$ m/s or $\dstb_r= 11$ m/s. These conditions correspond to \textit{moderate breeze} and \textit{strong breeze} respectively on the Beaufort scale \cite{Windscale}. The scheduled times of arrival $\sta_i$ for all vehicles are chosen to be all $0$ for a high UAV density condition. For medium and low density conditions, we separated the $\sta_i$ for the vehicles by $5$ s and $10$ s respectively, with the latest $\sta_i$ being 0. Note that we have used same dynamics and input bounds across all vehicles for clarity of illustration; however, SPP can easily handle more general systems of the form in which the vehicles have different control bounds, $\sta_i$ and dynamics.

The goal of the vehicles is to reach their destinations while avoiding a collision with the other vehicles or the static obstacles. The joint state space of this 50-vehicle system is 150-dimensional (150D), making the joint path planning and collision avoidance problem intractable for direct analysis. Therefore, we assign a priority ordering to vehicles and solve the path planning problem sequentially. For this simulation, we assign a random priority order to fifty vehicles and use RTT algorithm to compute the trajectories of the vehicles. 
% !TEX root = ../../SPP_IoTjournal.tex
\subsection{Results for $6$ m/s wind and no arrival time separation\label{sec:city_simResults}}
As per Algorithm \ref{alg:rtt}, we start with $\veh_1$ and sequentially compute the optimal control policy $\tracklaw_j$ and the latest departure time $\ldt_j$ for each vehicle. To compute the error bound $\errorbound_j$ on the tracking error of vehicle $j$, we choose $R_{\text{EB}} = 5$ m and use the reduced control authority $\cset^p_j = \{(v_{r,j}, \omega_{r,j}): 11 \text{ m/s} \leq v_{r,j} \leq 13 \text{ m/s}, |\omega_{r,j}| \leq 1.2 \text{ rad/s}\}$. Given dynamics in \eqref{eq:dyn_i}, the error dynamics between $\veh_j$ and the virtual evader are given by \cite{Mitchell05}:
\begin{equation}
\label{eq:reldyn}
\begin{aligned}
\dot{e}_{x,j} &= v_{j} \cos(e_{\theta,j}) - v_{r,j} + \omega_{r,j}{e}_{y,j} + d_{x,j}\\
\dot{e}_{y,j} &= v_{r,j}\sin(e_{\theta,j}) - \omega_{r,j}{e}_{x,j} + d_{y,j}\\
\dot{e}_{\theta,j} &= \omega_{j} - \omega_{r,j},
\end{aligned}
\end{equation}    
where $e_j = ({e}_{x,j}, {e}_{y,j}, {e}_{\theta,j})$ is the tracking error in the three states of $\veh_j$. Given relative dynamics, we compute the BRS $\brs^{\text{EB}}$ using \eqref{eqn:errBound} and evaluate the infinite-horizon control invariant set $\disckernel_j$. For all the BRS computations in this simulation, we use Level Set Toolbox \cite{Mitchell07b}. In presence of moderate winds, the obtained tracking error bound is $5$ m. This means that given any trajectory (which is a sequence of states over time) of vehicle, winds can at most cause a deviation of $5$ m from this trajectory at all times. Consequently, the vehicle will be within a distance of $5$ m from the trajectory. Note that since all SPP vehicles have same dynamics, the error bound is also same for all vehicles. 

The optimal control policy for $\veh_j$ is thus given by $\tracklaw_j(e_j)$ in \eqref{eq:robust_tracking_law}. However, to compute the relative state $e_j$, the nominal trajectory $\state_{r,j}$ for $\veh_j$ is required. Using $\disckernel_j$, we compute the obstacles induced by the higher-priority vehicles for $\veh_j$ and the obstacle induced by $\veh_j$ for the lower-priority vehicles. These obstacles are given by \eqref{eqn:rttAugObs} and \eqref{eqn:rttObs} respectively. Note that since disturbance directly impacts the computation of tracking error bound, these obstacles also grow as disturbance magnitude increases. We will illustrate the effect of disturbance magnitude on the trajectories of vehicles in \ref{sec:city_sim}-\ref{sec:city_distbEffect}.

The nominal trajectory can now be obtained by computing $\brs_j^\text{rtt}$ in \eqref{eq:rttBRS} and executing the corresponding control policy $\ctrl_j^\text{rtt}$ in \eqref{eq:RTTOptCtrl}, starting from the initial state $\state^0_j$. Finally, the latest departure time $\ldt_j$ is given by $\arg \sup_t \state^0_j \in \brs_j^\text{rtt}(t, \sta_j)$. It is important to note that since the BRS $\brs_j^\text{rtt}$ is computed backwards starting from the scheduled time of arrival $\sta_j$, (a) the latest departure time $\ldt_j$ directly depends on and varies with $\sta_j$ and (b) it directly impacts the obstacles that $\veh_j$ needs to avoid in its path towards its destination. We will illustrate the effect of the scheduled time of arrival on the trajectories of vehicles also in \ref{sec:city_sim}-\ref{sec:city_distbEffect}.

\begin{figure*}[!htb]
 \centering
\begin{subfigure}{0.5\textwidth}
  \includegraphics[width=\columnwidth]{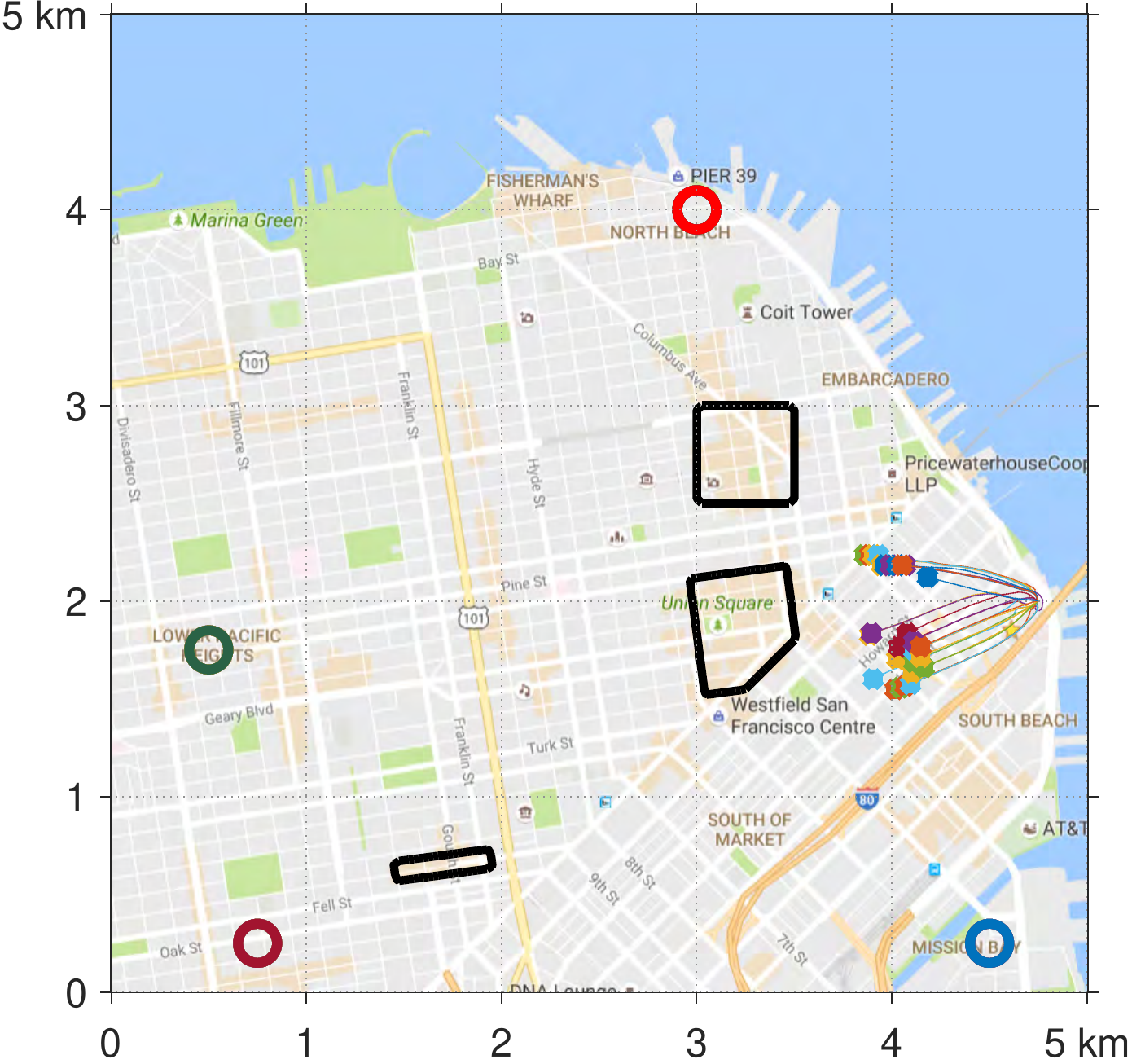}
  \subcaption{}
  \label{fig:sf_d6sep0_s1}
\end{subfigure}%
\begin{subfigure}{0.5\textwidth}
  \includegraphics[width=\columnwidth]{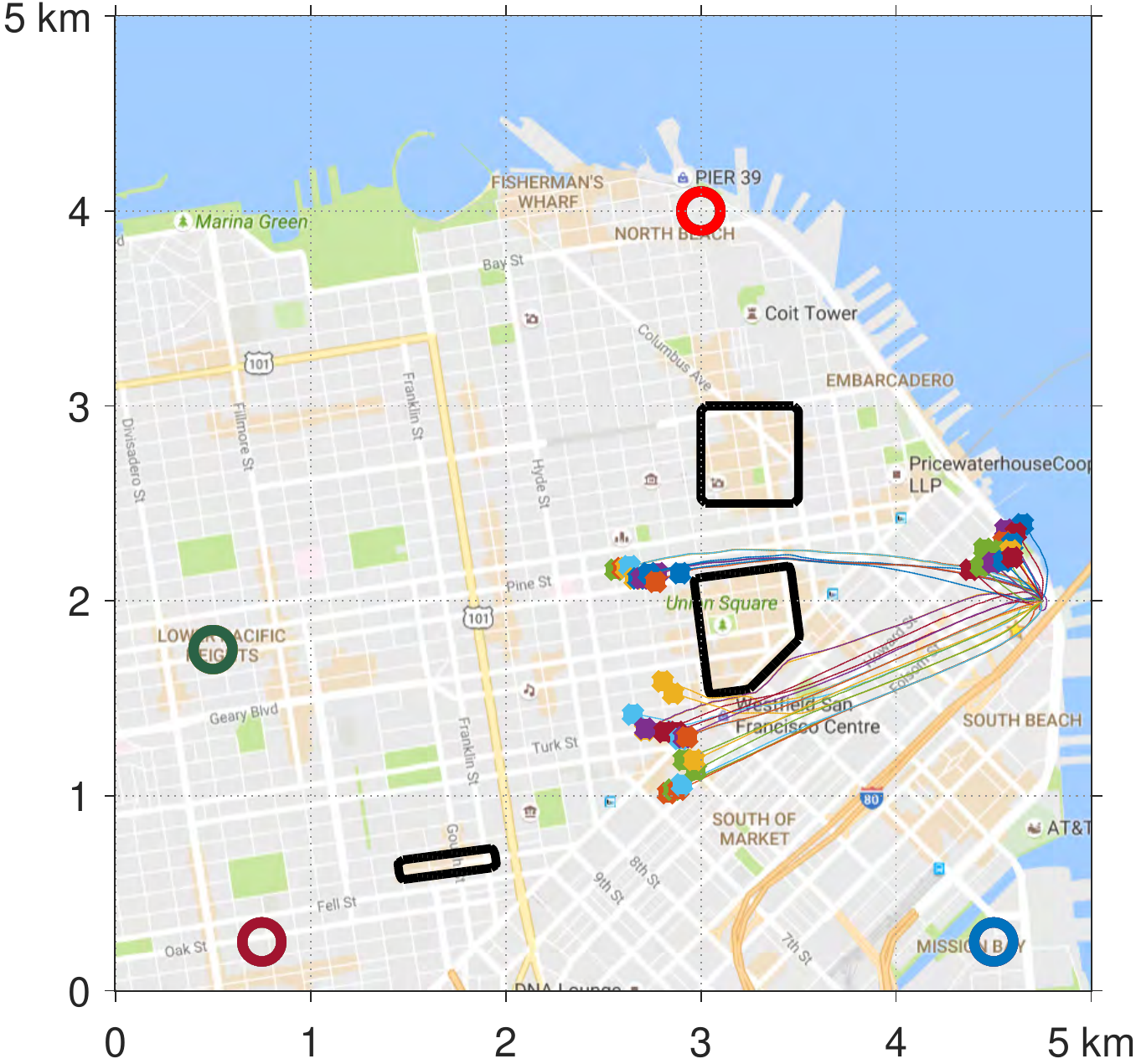}
  \subcaption{}
  \label{fig:sf_d6sep0_s2}
\end{subfigure}%

\begin{subfigure}{0.5\textwidth}
  \includegraphics[width=\columnwidth]{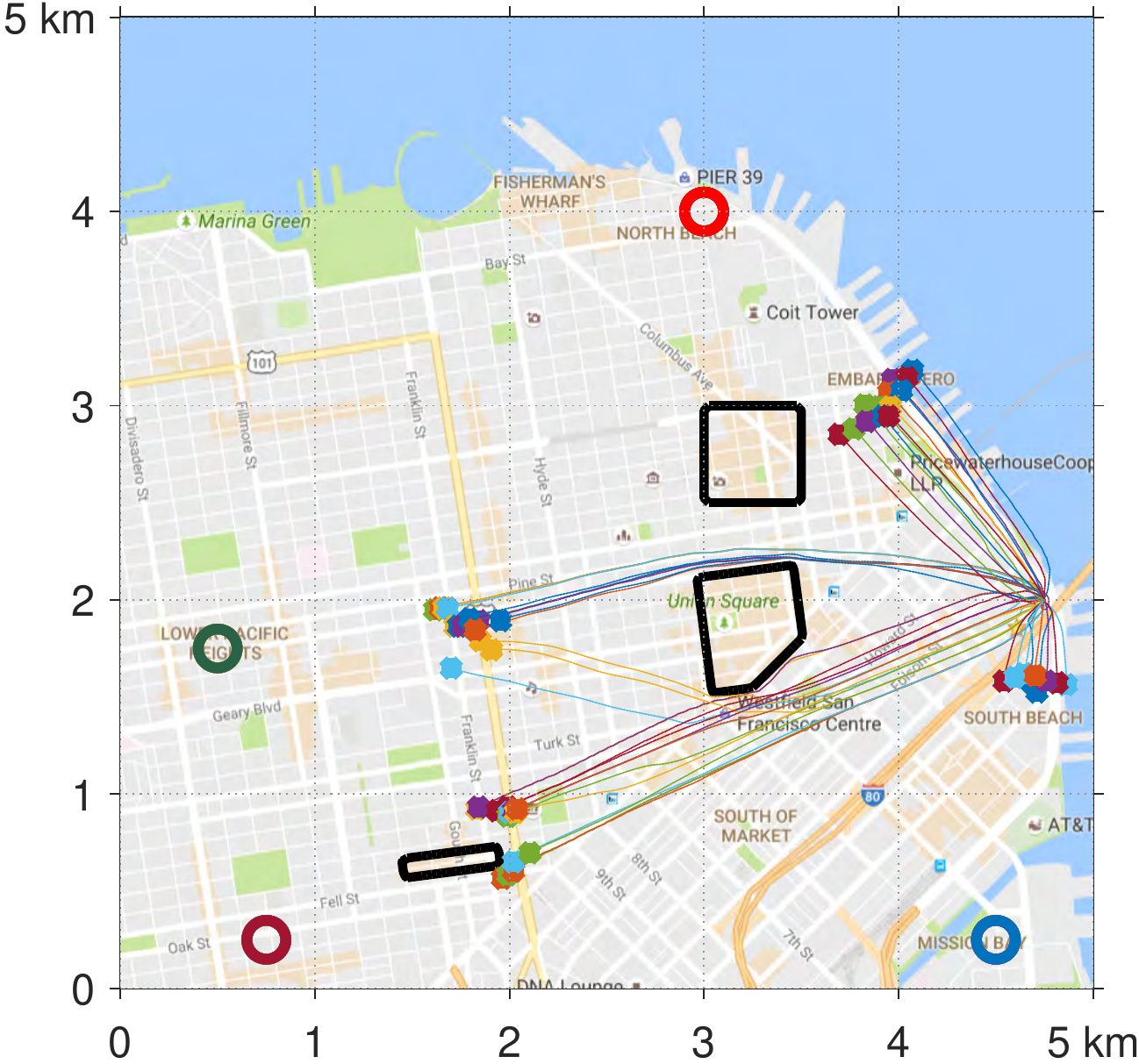}
  \subcaption{}
  \label{fig:sf_d6sep0_s3}
\end{subfigure}%
\begin{subfigure}{0.5\textwidth}
  \includegraphics[width=\columnwidth]{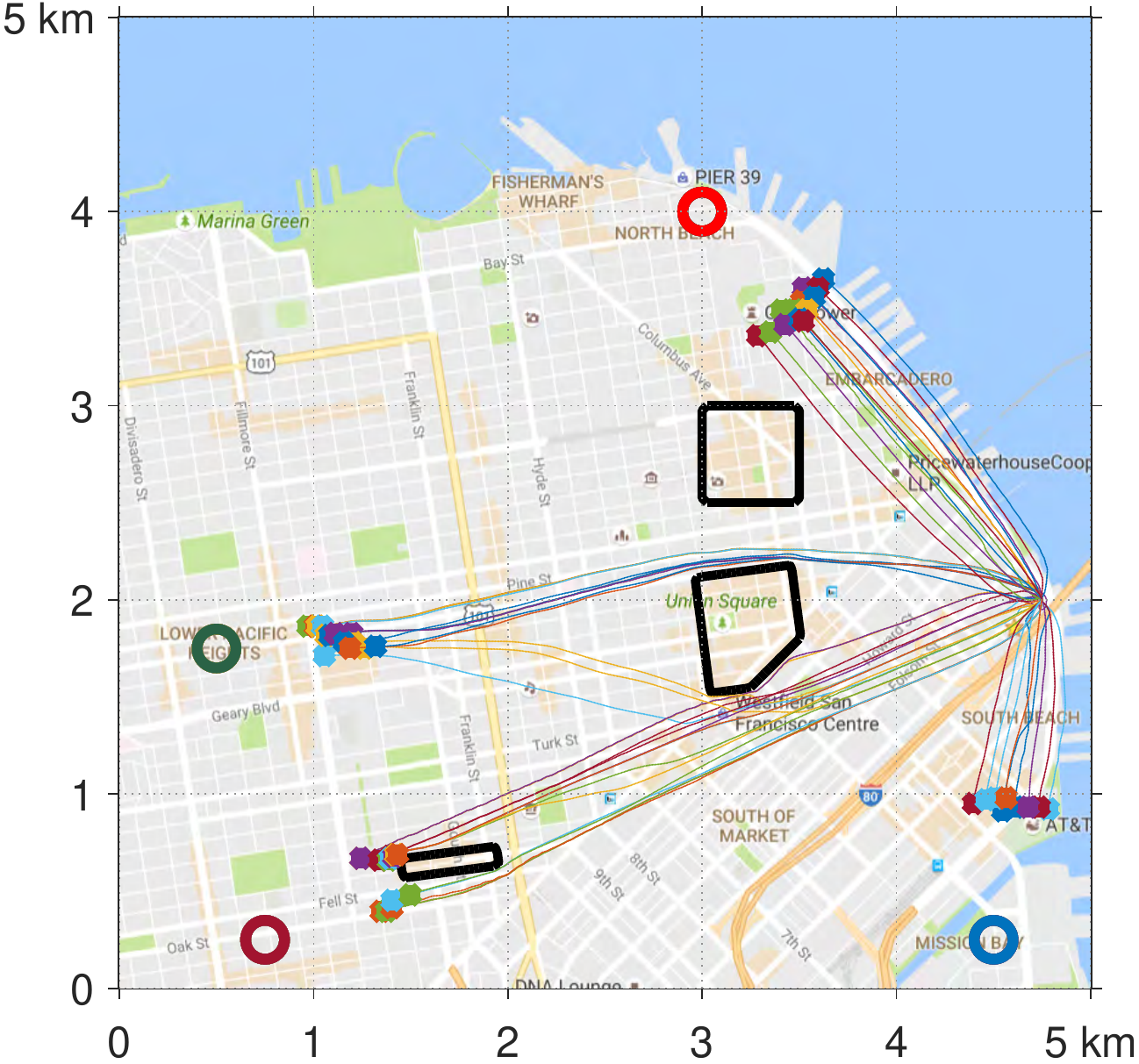}
  \subcaption{}
  \label{fig:sf_d6sep0_s4}
\end{subfigure}%
\caption{Snapshots of vehicle trajectories at different times for uniform disturbance with $d_{r} = 6$ m/s. The vehicles remain clear of all static obstacles despite the disturbance in the dynamics.}
\label{fig:trajectories_sf_snapshots}
\end{figure*}

The resulting trajectories of vehicles for $d_{r} = 6$ m/s and $\sta_j = 0 ~ \forall j$ at different times are shown in Figure \ref{fig:trajectories_sf_snapshots}. As evident from the figures, the vehicles remain clear of all the static obstacles (the black contours) and make steady progress towards reaching their destinations. The vehicles whose destinations are relatively closer need less time to travel to their destinations and depart later. Note that the shown trajectories are simulated under uniformly random disturbance (i.e., for every vehicle, the disturbance is uniformly sampled from a circle of radius $d_{r} = 6$ m/s at each time step), but the SPP algorithm guarantees safety and reactivity despite the worst case disturbance, as we show later in this section.

\begin{figure}[h]
  \centering
  \includegraphics[width=\columnwidth]{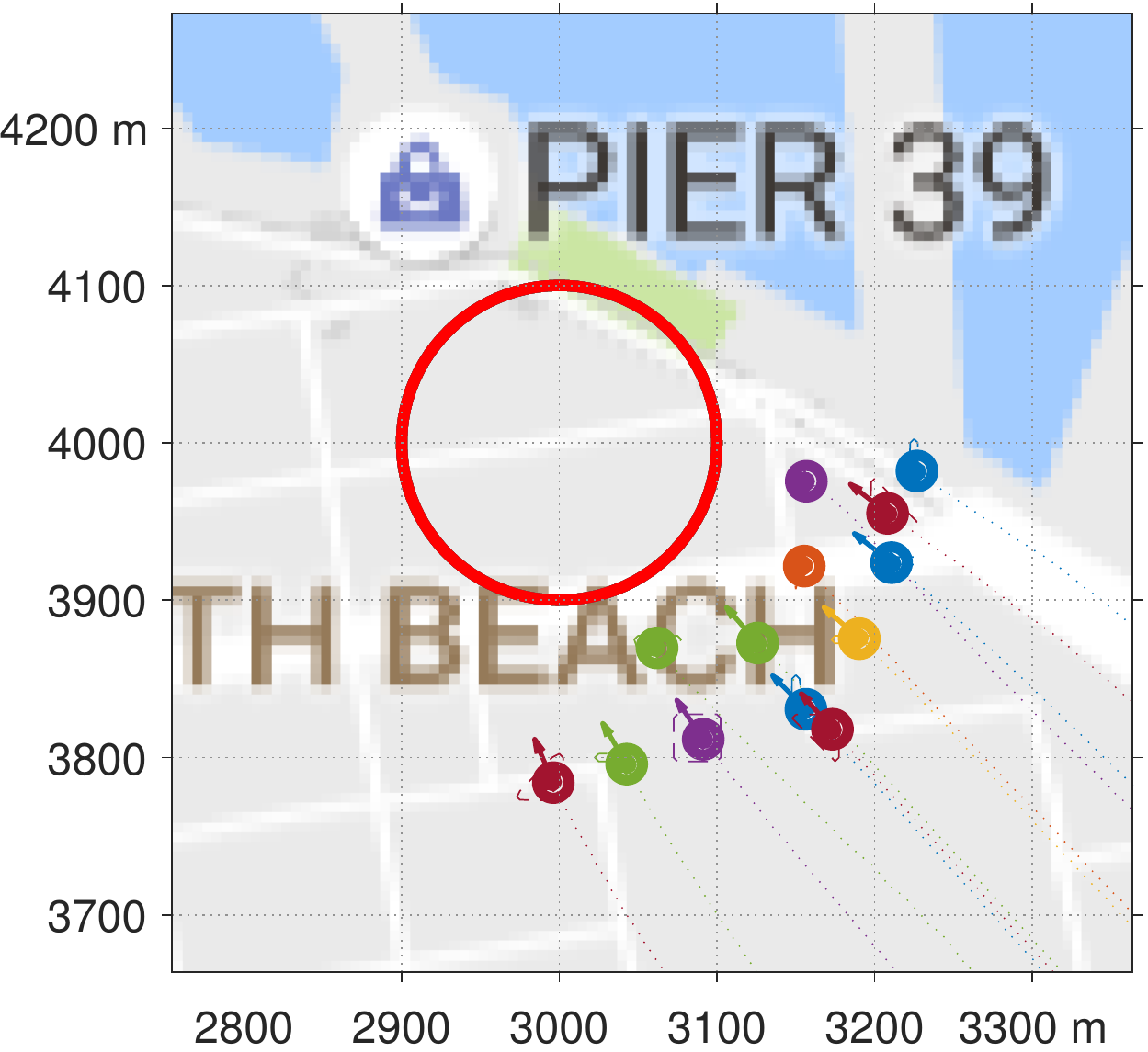}
  \caption{Zoomed-in version of vehicle trajectories near the red target in Figure \ref{fig:sf_d6sep0}. The SPP algorithm ensures that the vehicles are outside each other's danger zones (circles around the vehicles).} 
  \label{fig:sf_d6sep0_zoomed}
\end{figure}

The full trajectories of vehicles are shown in Figure \ref{fig:sf_d6sep0}. All vehicles reach their respective destinations. A zoomed-in version of Figure \ref{fig:sf_d6sep0} near the red-target (Figure \ref{fig:sf_d6sep0_zoomed}) illustrates that vehicles are also outside each other's danger zones (circles around the vehicles) as required. 
It is interesting to note that the vehicles going to the same destination take different trajectories. This is because all vehicles have same scheduled time of arrival, and hence the lower-priority vehicles do not have the flexibility to wait for the higher-priority vehicles. In order to ensure that they reach their destinations on time, they take alternative trajectories to their destinations, forming different ``traffic lanes''. Thus, the vehicles' trajectories obtained in this case are predominately \textit{state-separated} trajectories, i.e., they follow different state trajectories but at the same time. 

\begin{figure*}[!htb]
  \centering
  \begin{subfigure}{0.5\textwidth}
    \includegraphics[width=\columnwidth]{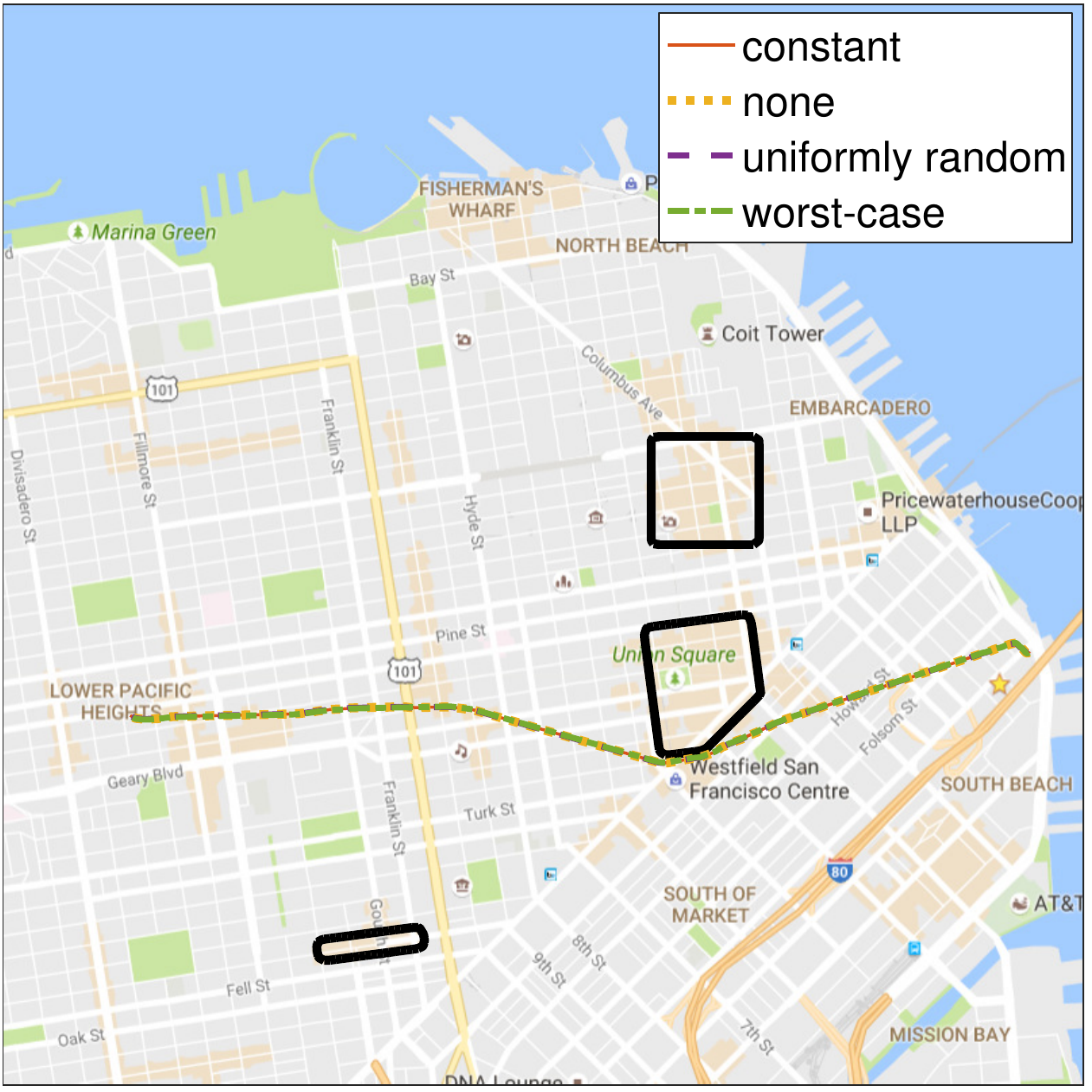}
    \subcaption{}
    \label{fig:dstb_trajs_s1}
  \end{subfigure}%
  \begin{subfigure}{0.5\textwidth}
    \includegraphics[width=\columnwidth]{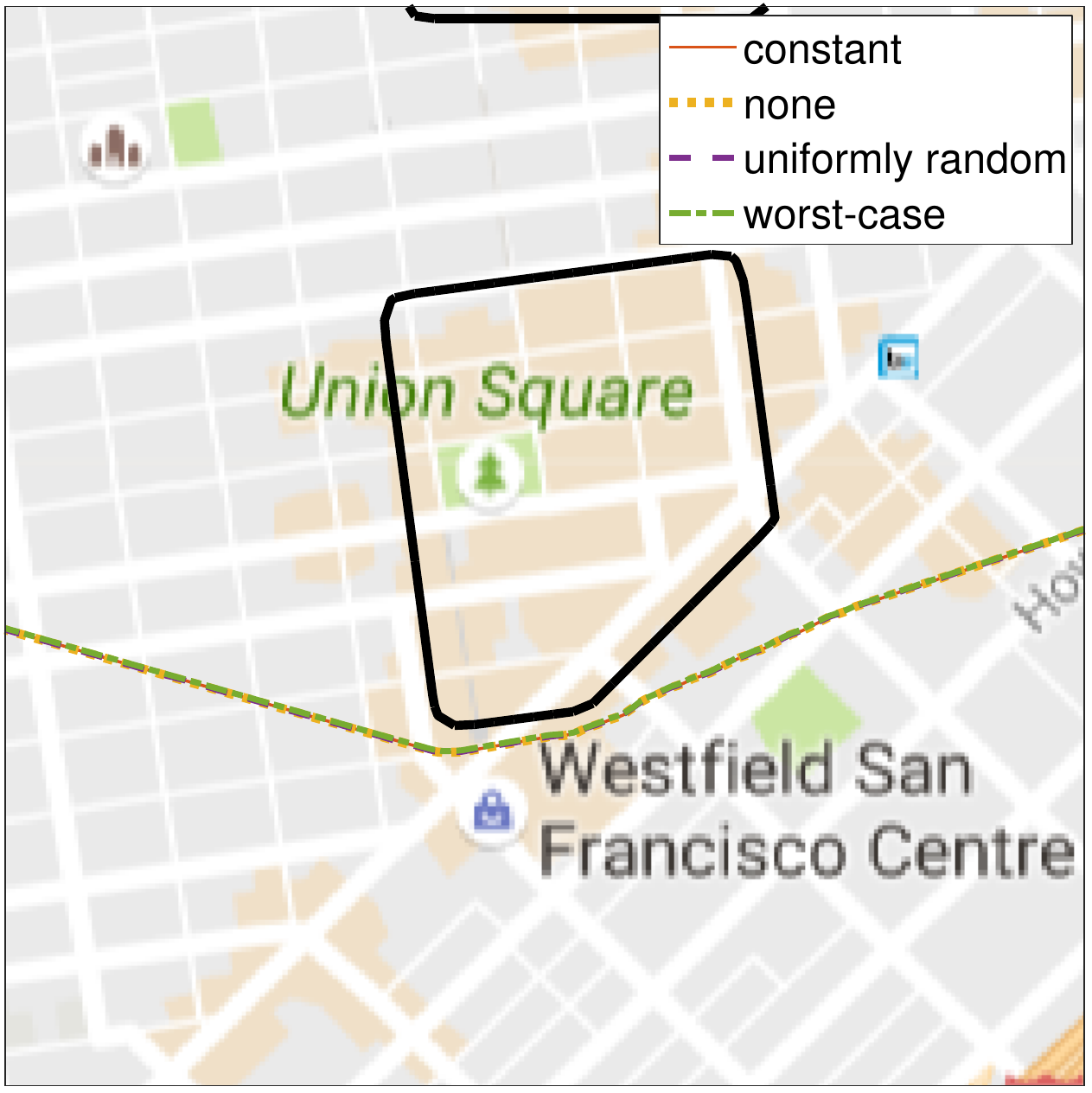}
    \subcaption{}
    \label{fig:dstb_trajs_s2}
  \end{subfigure}%

  \caption{Trajectories of $\veh_{17}$ under different types of bounded disturbances -- constant wind from the west, no wind, uniformly random wind, and worse-case wind. The right plot zooms in on the beginning of the left plot.}
  \label{fig:dstb_trajs}
\end{figure*}

Although the plotted trajectories in Figure \ref{fig:sf_d6sep0} are for a particular realization of the (uniformly random) disturbances, the SPP algorithm guarantees obstacle avoidance regardless of the realized disturbance. To illustrate this, we plot the trajectories of a particular vehicle ($\veh_17$) for three different disturbance realizations -- uniform, worst-case, and constant wind from the west direction -- in Figure \ref{fig:dstb_trajs}. One can see that the trajectories are nearly indistinguishable even in the zoomed-in plot on the right. 

\begin{figure*}[!htb]
  \centering
  \begin{subfigure}{0.5\textwidth}
    \includegraphics[width=\columnwidth]{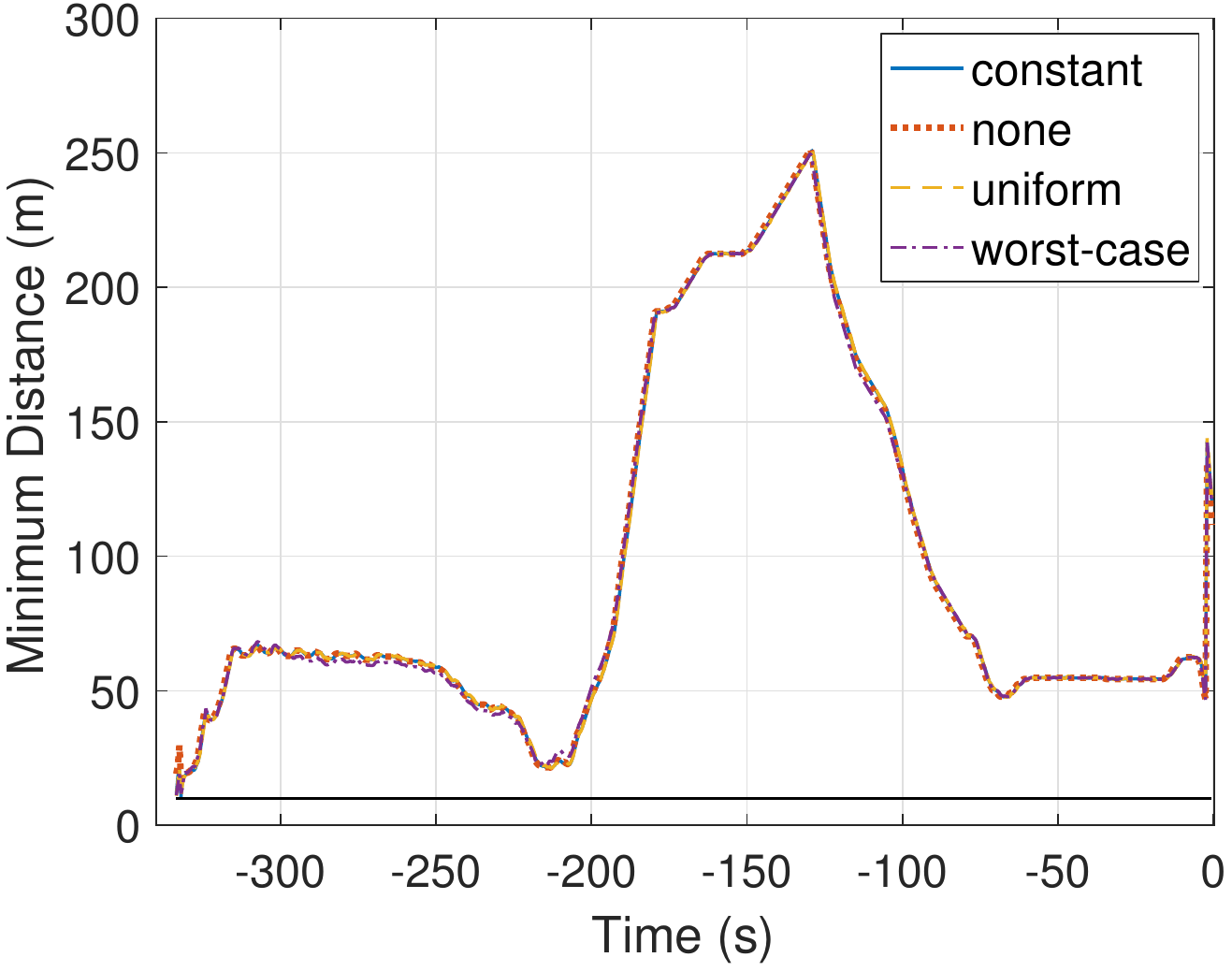}
    \subcaption{}
    \label{fig:min_dists_s1}
  \end{subfigure}%
  \begin{subfigure}{0.5\textwidth}
    \includegraphics[width=\columnwidth]{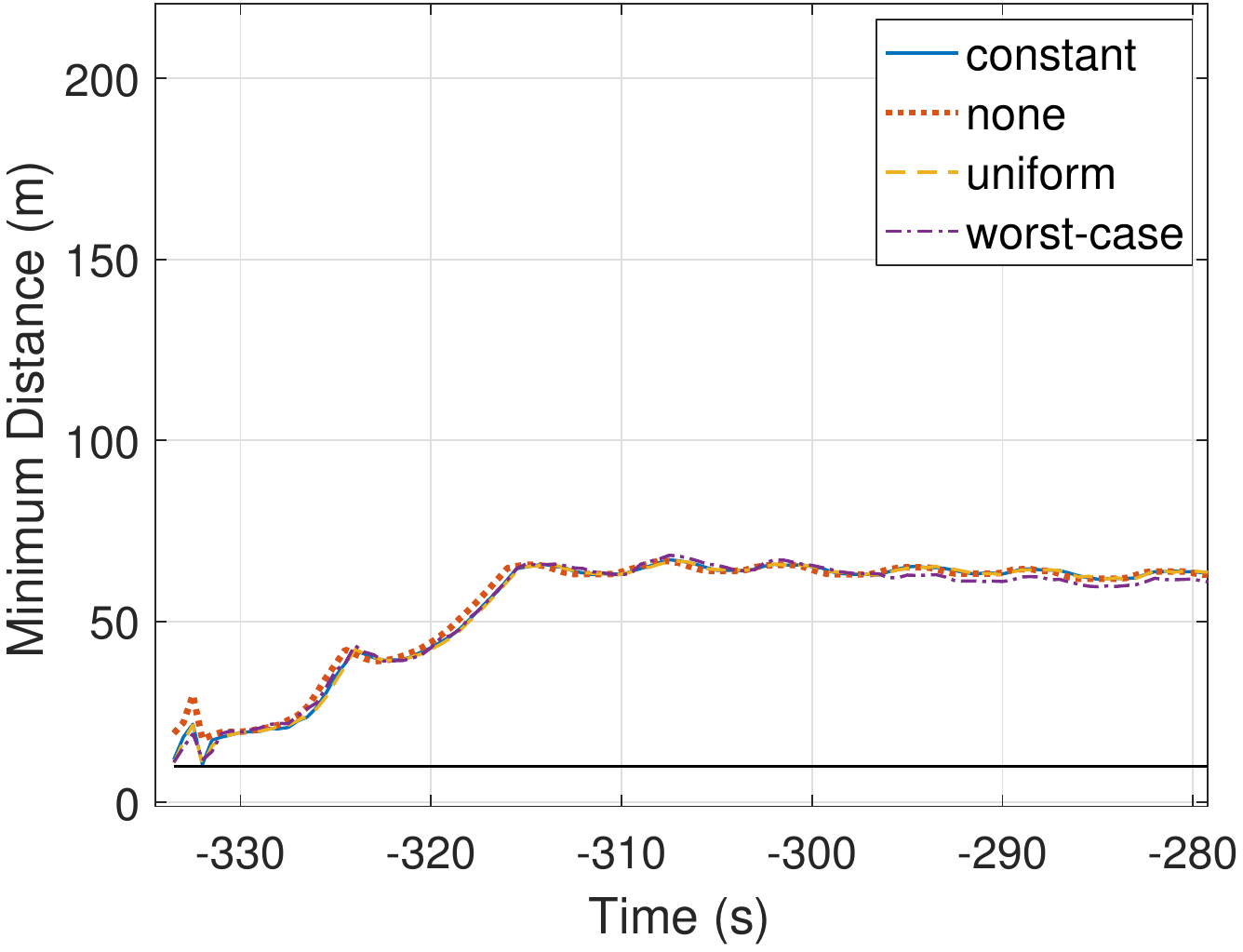}
    \subcaption{}
    \label{fig:min_dists_s2}
  \end{subfigure}%
  
  \caption{Minimum distance to other vehicles for $\veh_{17}$ under different types of bounded disturbances -- constant wind from the west, no wind, uniformly random wind, and worse-case wind. The right plot zooms in on the beginning of the left plot.}
  \label{fig:min_dists}
\end{figure*}

Figure \ref{fig:min_dists} shows the minimum distance between $\veh_{17}$ and other vehicles. One can see that this distance is never below $10$ m, which is minimum required separation between the vehicles. The separation distance is small in the beginning because the vehicles need to depart as soon as possible while maintaining this minimum required separation. 

\begin{figure*}[!htb]
  \centering
  \begin{subfigure}{0.5\textwidth}
    \includegraphics[width=\columnwidth]{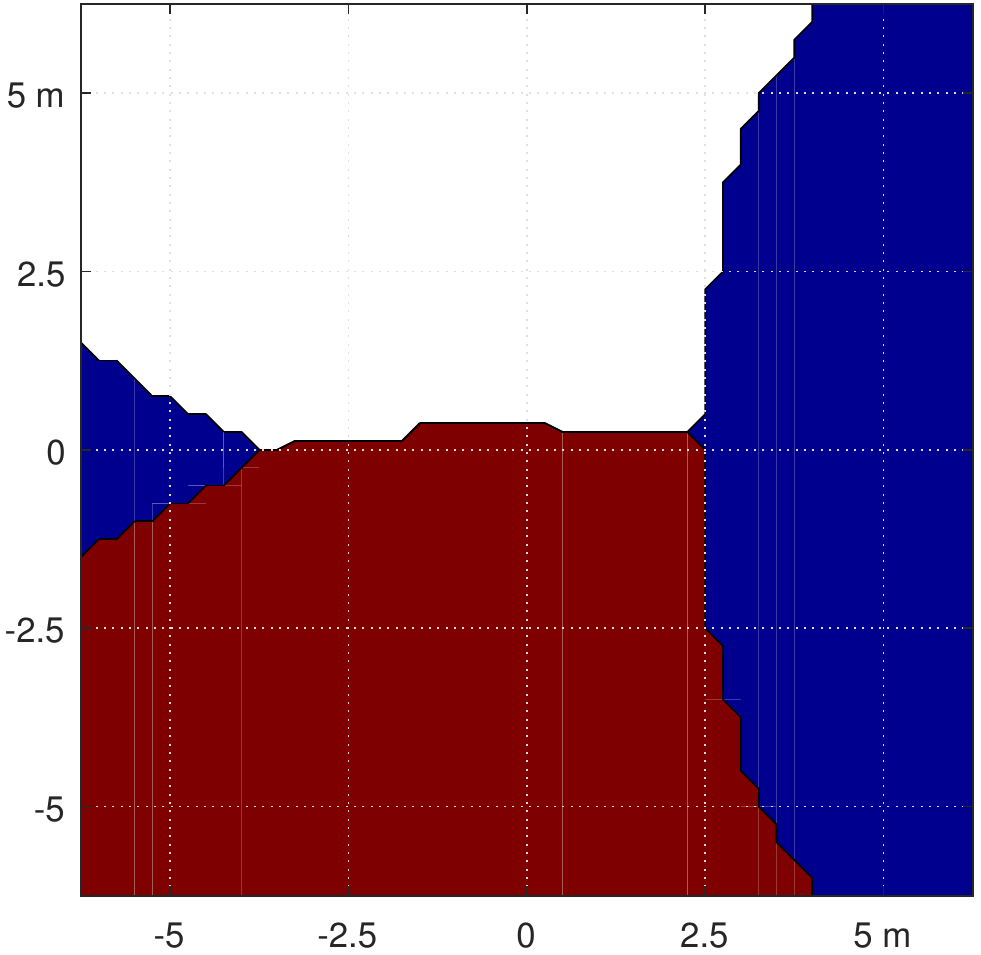}
    \subcaption{Turn rate control as a function of tracking error. White: turn right; red: turn left; blue: go straight.}
    \label{fig:Wcontrol}
  \end{subfigure}%
  \begin{subfigure}{0.5\textwidth}
    \includegraphics[width=\columnwidth]{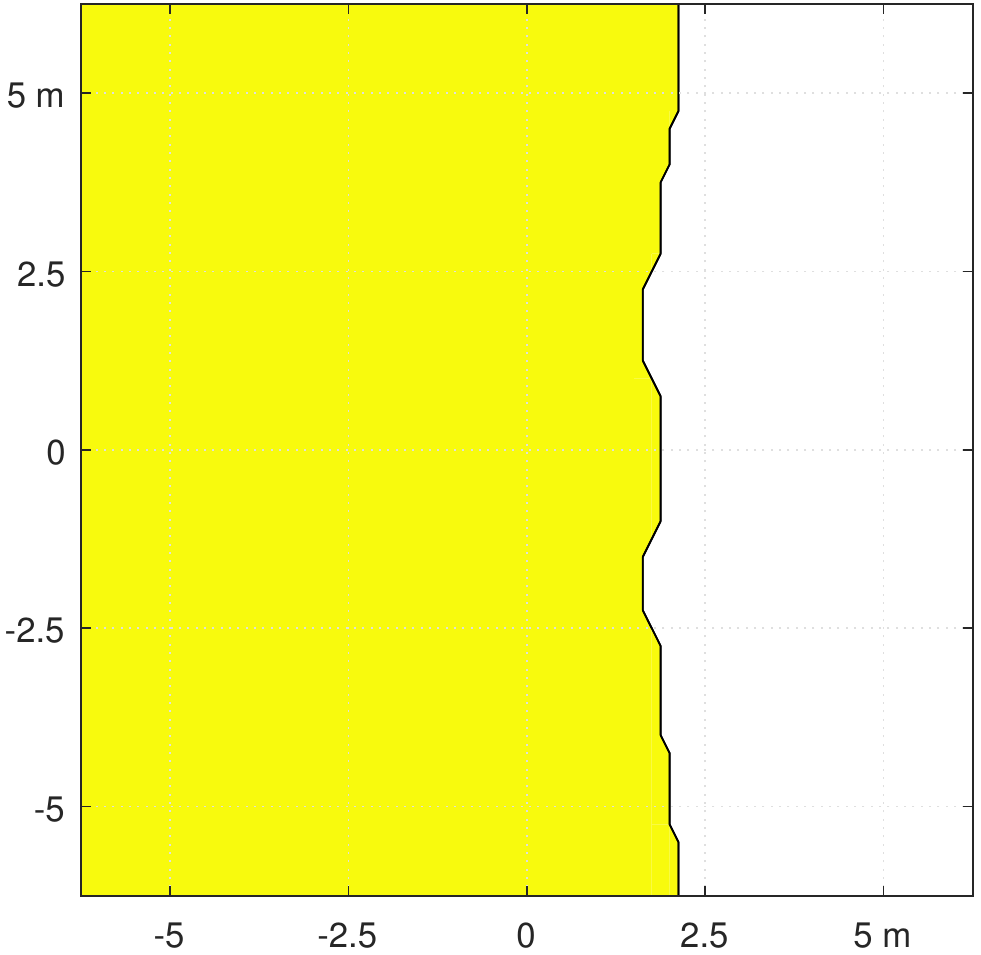}
    \subcaption{Linear speed control as a function of tracking error. Yellow: high speed; white: low speed.}
    \label{fig:Vcontrol}
  \end{subfigure}%
  
  \caption{Optimal tracking control law for each vehicle for a heading of zero degrees (facing towards the right).}
  \label{fig:reactivity}
\end{figure*}

Figure \ref{fig:reactivity} shows the optimal control action in terms of the turn rate and the linear speed of each vehicle, demonstrating that the tracking control law reacts to tracking error due to disturbances in an intuitive way. Both subplots show the control law when the vehicle is facing east. Figure \ref{fig:Wcontrol} indicates, for example, that generally when the tracking error is such that the vehicle is too far to the right compared to the nominal position, the optimal tracking control is to turn left. Figure \ref{fig:Vcontrol} indicates, for example, that when the vehicle is too far ahead of the nominal position, it should slow down. The optimal turn rate and linear speed controls ensure that the vehicle never deviates from the nominal trajectory for more than $5$ m.

The average trajectory computation time per vehicle is 2 s using a CUDA implementation of the level set toolbox on a desktop computer with a Core i7 5820K processor and two GeForce GTX Titan X graphics processing units. Note that all this computation is done offline and the resulting optimal policy $\tracklaw_j(e_j)$ is obtained as a lookup table. In real time, neither any computation nor any communication between vehicles is required. Only a lookup table query is required, and this can be performed very quickly in real time. This illustrates the capability of SPP as a provably safe path planning algorithm for large multi-vehicle systems. Without CUDA, the computation time per vehicle is 33 s using a C++ implementation of the level set toolbox, and 200 s using a Matlab implementation.
% !TEX root = ../../SPP_IoTjournal.tex
\subsection{Effect of Disturbance and Scheduled Time of Arrival \label{sec:city_distbEffect}}
In this section, we illustrate how the disturbance bound $d_r$ in \eqref{eq:dyn_i} and the relative $\sta$s of vehicles affect the vehicle trajectories. For this purpose, we simulate the SPP algorithm for four additional scenarios:
\begin{itemize}
\item Case-0: $d_r = 6$ m/s, $\sta_i = 0 ~\forall i$
\item Case-1: $d_r = 11$ m/s, $\sta_i = 0 ~\forall i$
\item Case-2: $d_r = 6$ m/s, $\sta_i = 5(i-1) ~\forall i$
\item Case-3: $d_r = 11$ m/s, $\sta_i = 5(i-1) ~\forall i$
\item Case-4: $d_r = 11$ m/s, $\sta_i = 10(i-1) ~\forall i$
\end{itemize}
The interpretation $\sta_i = 5(i-1)$ is that the scheduled time of arrival of any two consecutive vehicles is separated by 5 s, which represents a medium vehicle density scenario; a separation of 10 s represents a low vehicle density scenario. $d_r = 6$ m/s and $d_r = 11$ m/s correspond to the moderate breeze and strong breeze respectively on Beaufort wind force scale \cite{Windscale}. 

\begin{figure*}[!htb]
 \centering
\begin{subfigure}{0.5\textwidth}
  \includegraphics[width=\columnwidth]{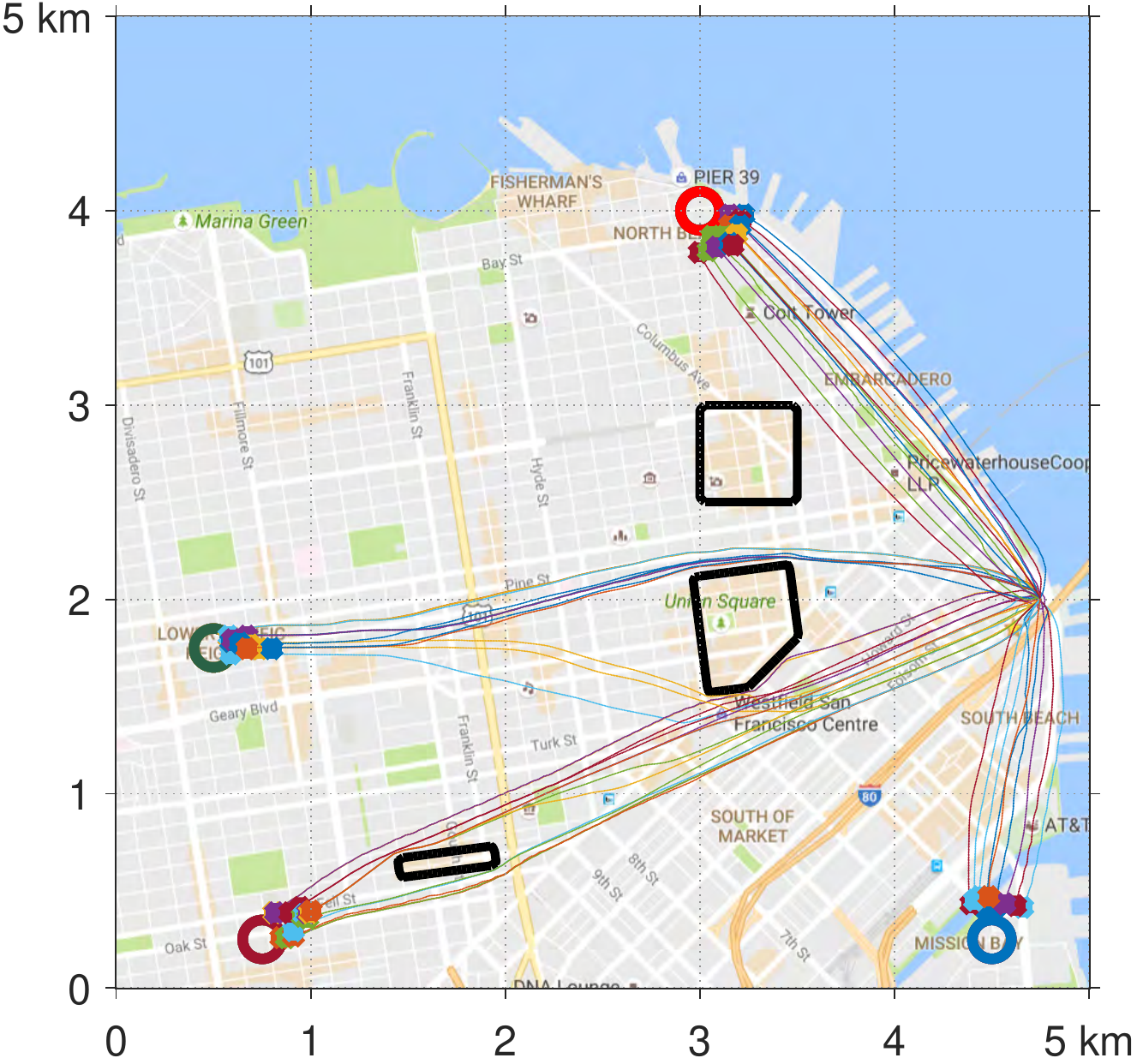}
  \subcaption{Case-0: $d_r = 6m/s$, $\sta_i = 0$}
  \label{fig:sf_d6sep0}
\end{subfigure}%
\begin{subfigure}{0.5\textwidth}
  \includegraphics[width=\columnwidth]{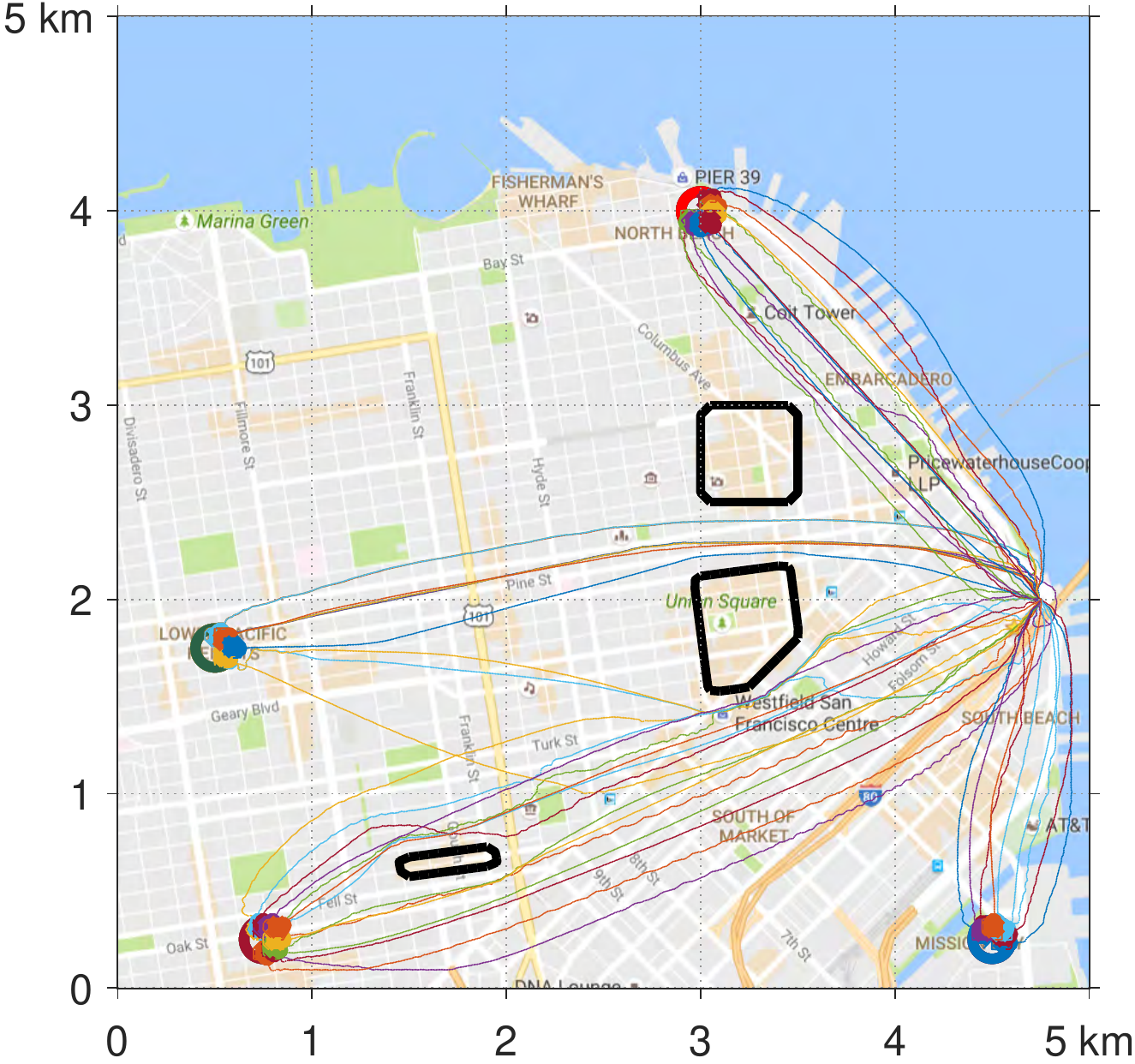}
  \subcaption{Case-1: $d_r = 11m/s$, $\sta_i = 0$}
  \label{fig:sf_d11sep0}
\end{subfigure}%

\begin{subfigure}{0.5\textwidth}
  \includegraphics[width=\columnwidth]{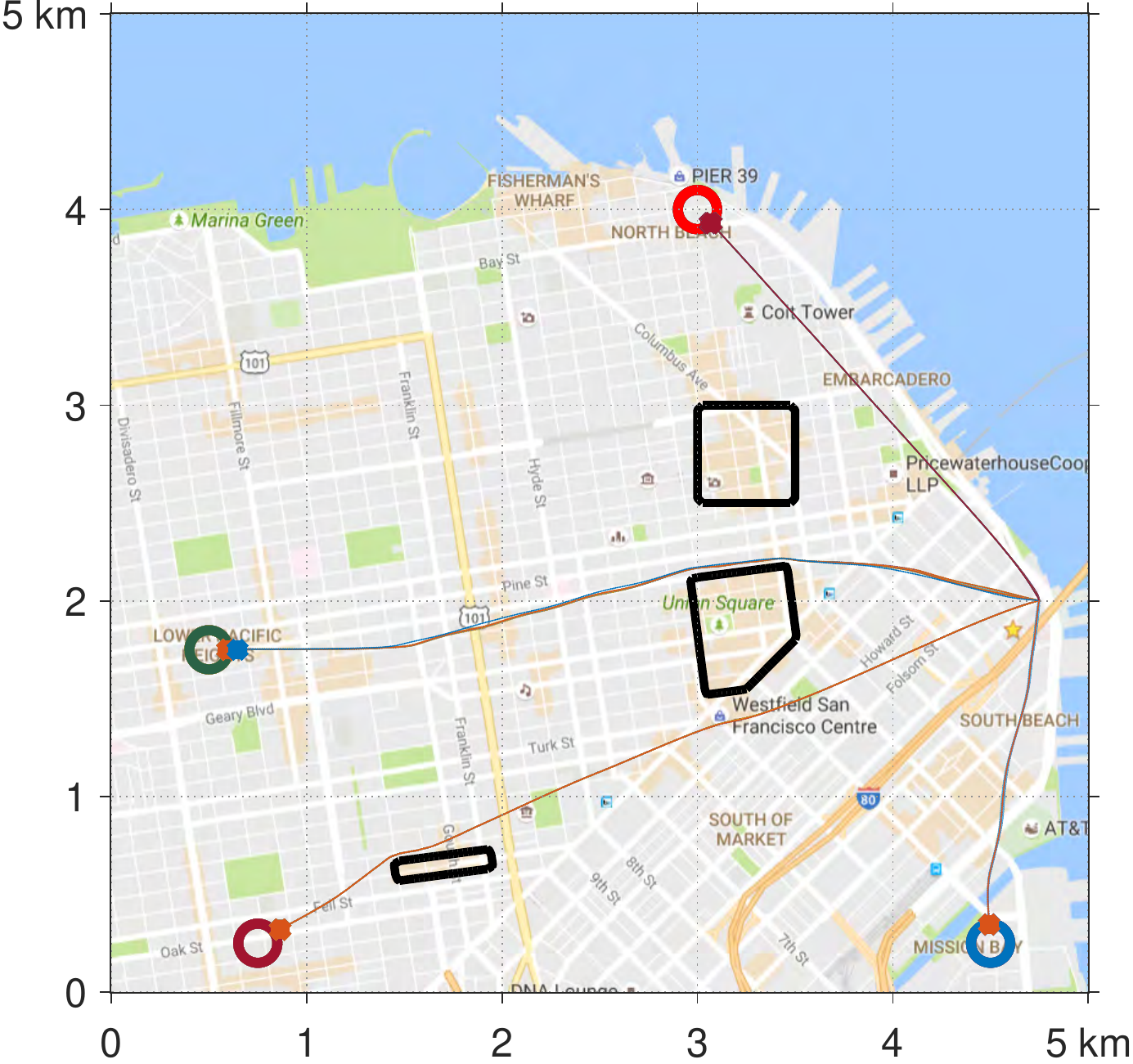}
  \subcaption{Case-2: $d_r = 6m/s$, $\sta_i = 5(i-1)$}
  \label{fig:sf_d6sep5}
\end{subfigure}%
\begin{subfigure}{0.5\textwidth}
  \includegraphics[width=\columnwidth]{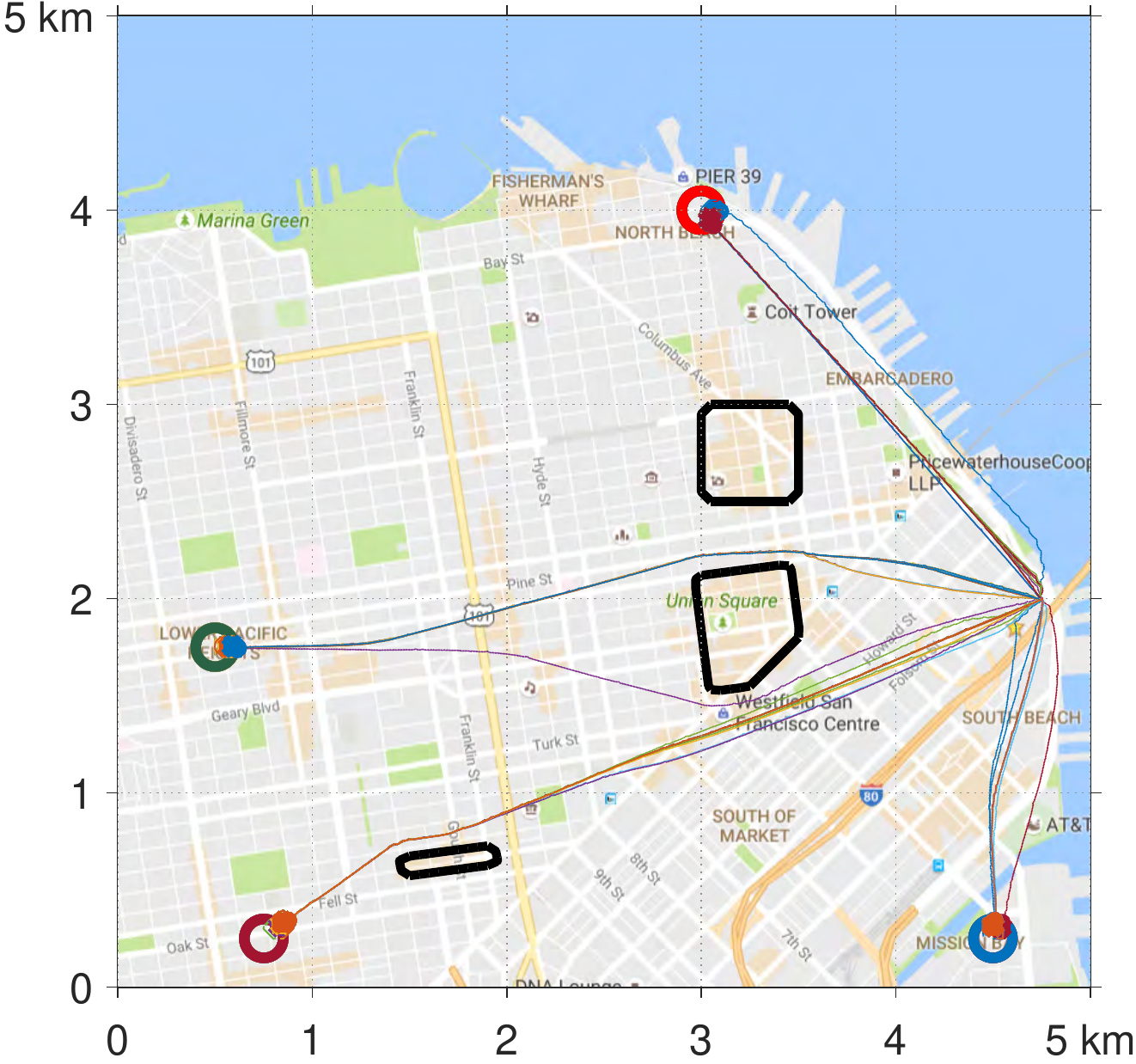}
  \subcaption{Case-3: $d_r = 11m/s$, $\sta_i = 5(i-1)$}
  \label{fig:sf_d11sep5}
\end{subfigure}%
\caption{Effect of the disturbance magnitude and the scheduled times of arrival on vehicle trajectories. All trajectories are simulated under uniformly random disturbance. The relative separation in the scheduled times of arrival of vehicles determines the number of lanes between a pair of origin and destination, and more and more tarjectories become time-separated as this relative separation increases. The disturbance magnitude determines the relative separation between different lanes, and more and more tarjectories become state-separated as the disturbance increases. }
\label{fig:trajectories_sf}
\end{figure*}

Intuitively, as $\dstb_r$ increases, it is harder for a vehicle to closely track a particular nominal trajectory, which results in a higher tracking error bound. As mentioned previously, with a $6$ m/s wind speed, the tracking error bound is 5 m; however, with an $11$ m/s wind speed, the tracking error bound becomes 35 m. Thus, the vehicles need to be separated more from each other in space to ensure that they do not enter each other's danger zones. This is also evident from comparing the results corresponding to Case-0 (Fig. \ref{fig:sf_d6sep0}) and Case-1 (Fig. \ref{fig:sf_d11sep0}). As the disturbance magnitude increases from $d_r = 6$ m/s (moderate breeze) to $d_r = 11$ m/s (strong breeze), the vehicles' trajectories get farther apart from each other. Since $\sta$ is same for all vehicles, the vehicles’ trajectories are still predominately \textit{state-separated} trajectories.

We next compare Case-0 and Case-2. The difference between these two cases is that vehicles have a 5 second separation in their schedule times of arrival in Case-2. When vehicles $\veh_i$ and $\veh_{j}$ ($j>i$) have same scheduled time of arrival and are going to the same destination, they are constrained to travel at the same time to make sure they reach the destination by the designated $\sta$. However, since $\veh_i$ is high-priority, it gets access to the optimal trajectory (in terms of the total time of travel to destination) and $\veh_{j}$ has to settle for a relatively sub-optimal trajectory. Thus, all vehicles going to a particular destination take different trajectories creating a ``band" of trajectories between the origin and the destination, as shown in Figure \ref{fig:sf_d6sep0}; the high-priority vehicles take a relatively straight path between the origin and the destination whereas the low-priority vehicles take a (relatively sub-optimal) curved path. If we think of an air highway between the origin and the destination, then vehicles take different lanes of that highway to reach the destination in Case-0. Thus, the trajectories of vehicles in this case are \textit{state-separated}. However, when $\sta_j > \sta_i$, then $\veh_j$ is not bound to travel at the same time as $\veh_i$; it can wait for $\veh_i$ to depart and take a shorter path later on. Thus, vehicles travel in a single (optimal) lane in this case, as shown in Figure \ref{fig:sf_d6sep5}. In other words, they take the same trajectory to the destination, but at different times. Thus, the trajectories of vehicles in this case are \textit{time-separated}. 

Note that the exact number of lanes depends on \textit{both} the disturbance and separation of scheduled times of arrival. As the disturbance increases, vehicles need to be separated more from each other to ensure safety. A larger arrival time difference between vehicles is also able to ensure this separation even if the vehicles were to take the same lane. As shown in Figure \ref{fig:sf_d11sep5}, a difference of 5 s in the $\sta$'s is not sufficient to achieve a single lane behavior for stronger 11 m/s wind conditions. However, the number of lanes is significantly less than that in Case-1 (Fig. \ref{fig:sf_d11sep0}). Finally, a separation of 10 s in $\sta$'s ensure that we get the single lane behavior even in the presence of 11 m/s winds, leading to \textit{time-separated} trajectories. Videos of the simulations can be found at https://youtu.be/1ocaBGZqSAE.

Overall, the relative magnitude of disturbance and scheduled times of arrival separation determines the number of lanes and type of trajectories that emerge out of the SPP algorithm. For a fixed disturbance magnitude, as the separation in the scheduled times of arrival of vehicles increases, the number of lanes between a pair of origin and destination decreases, and more and more trajectories become time-separated. On the other hand, for a fixed separation in the scheduled times of arrival of vehicles, as the disturbance magnitude increases, the number of lanes between a pair of origin and destination increases, and more and more trajectories become state-separated.

\begin{figure}[t]
  \centering
  \includegraphics[width=\columnwidth]{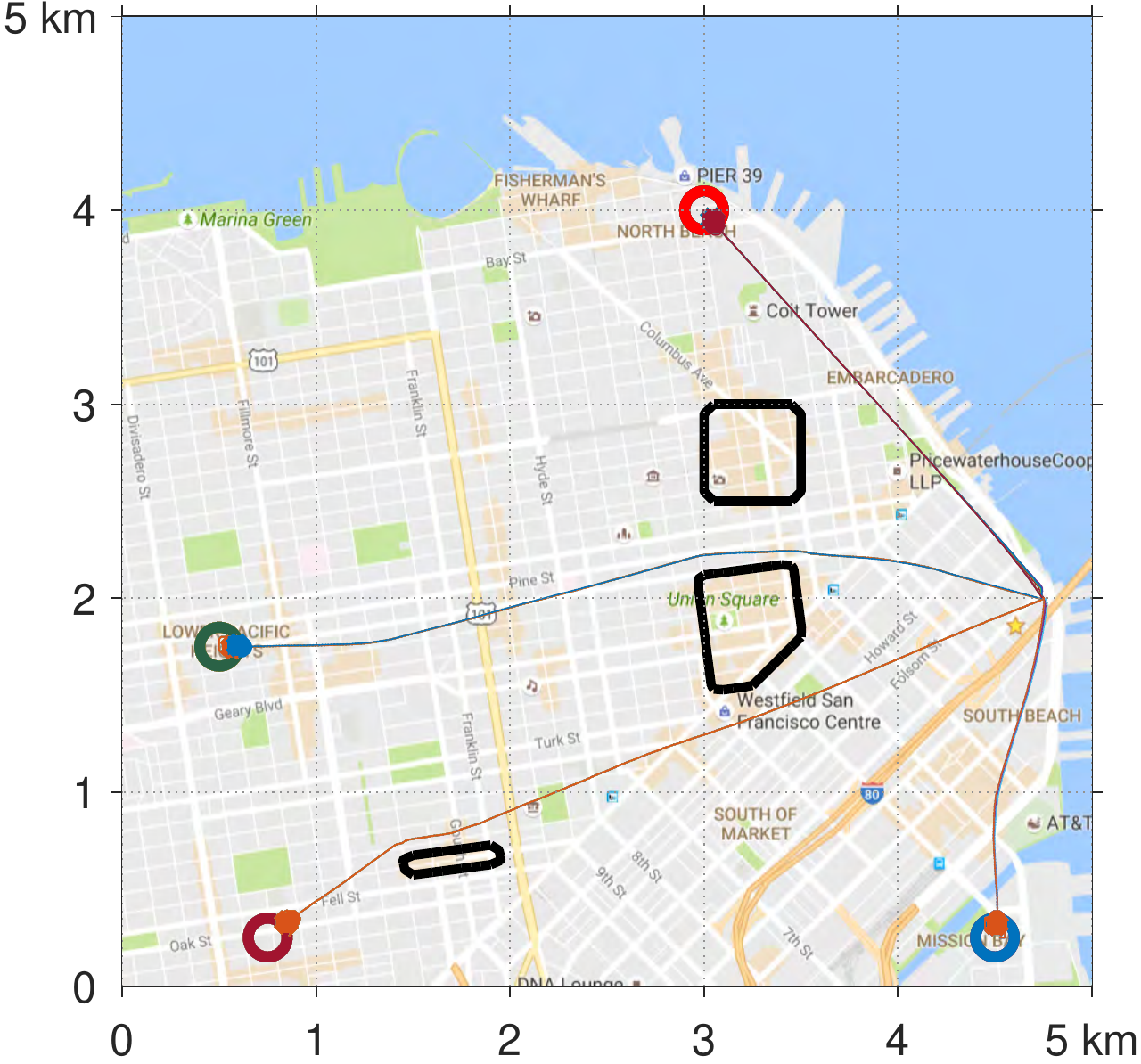}
  \caption{Vehicle trajectories for Case-4: $d_r = 11m/s$, $\sta_i = 10(i-1)$. Since different vehicles have different scheduled times of arrival, there is a single lane between every origin-destination pair.} 
  \label{fig:sf_d11sep10}
\end{figure}

% Bay Area level Simulations
% !TEX root = ../SPP_IoTjournal.tex
\section{Multi-city Environment Simulation \label{sec:bayArea_sim}}
We next use SPP algorithm to design trajectories for a 200-vehicle UAV system where UAVs are flying through a multi-city region.
% !TEX root = ../../SPP_IoTjournal.tex
\subsection{Setup \label{sec:bayArea_simSetup}}
\begin{figure}
  \centering
  \includegraphics[width=\columnwidth]{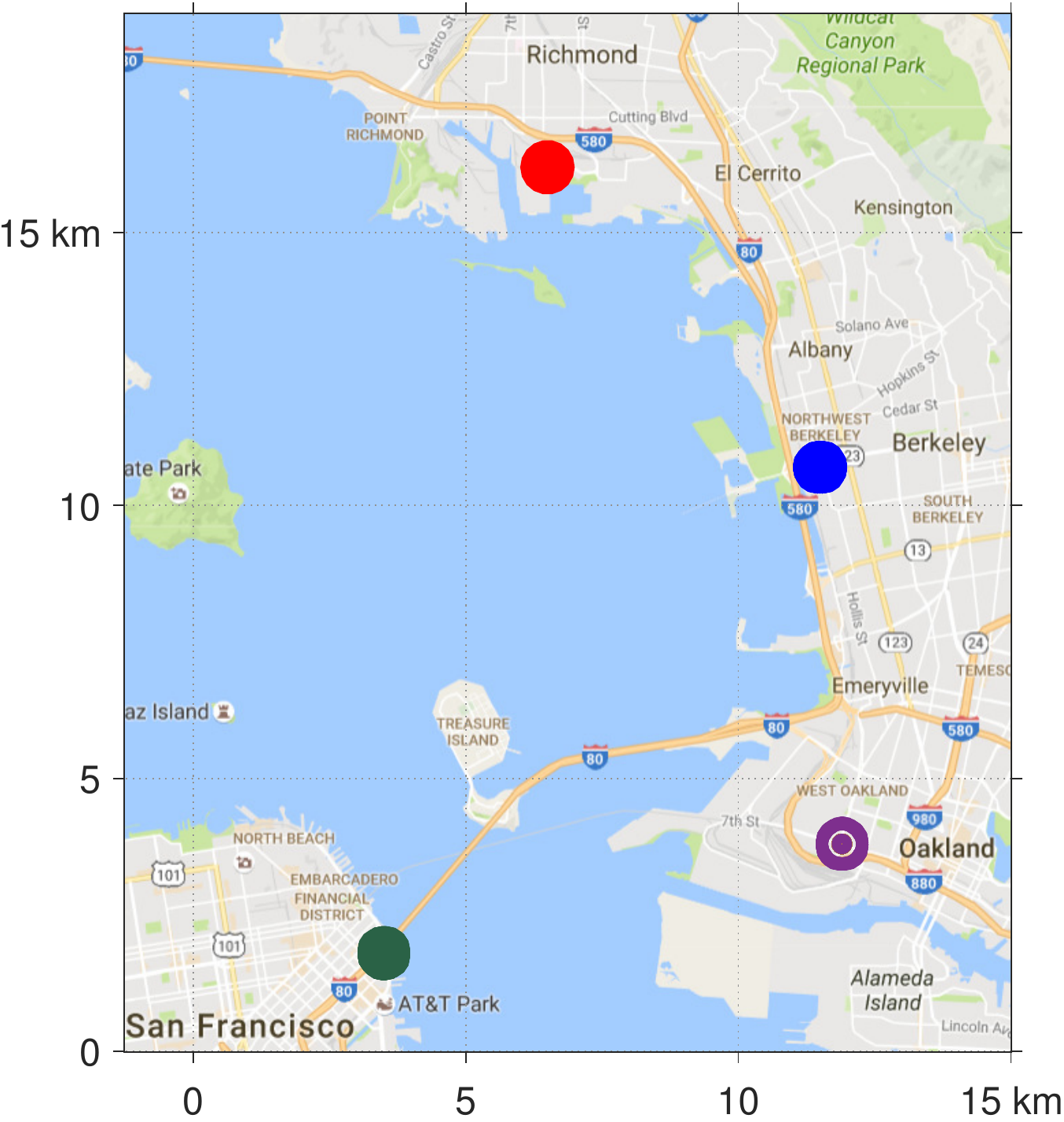}
  \caption{Multi-city simulation setup. A $300 km^2$ area of San Francisco Bay Area is used as the state-space for vehicles. SPP vehicles fly to and from the four cities indicated by the four circles. The simulations are performed under the strong winds condition with $d_{r} = 11 m/s$.}
  \label{fig:bayArea_setup}
\end{figure}

We grid the San Francisco Bay Area in California, US and use it as our state space, as shown in Figure \ref{fig:bayArea_setup}. We consider the UAVs flying to and from four cities: Richmond, Berkeley, Oakland, and San Francisco. The blue region in Fig. \ref{fig:bayArea_setup} represents bay. This environment is different from the city environment in Section \ref{sec:city_sim} in that now the UAVs need to fly for longer distances and through a high-density vehicle environment with strong winds, but have very few static obstacles like tall buildings.    

Each box in Figure \ref{fig:bayArea_setup} represents a $25$ km$^2$ area. The vehicles are flying to and from the four cities indicated by the four circles. The origin and the destination of each vehicle is chosen randomly from these four cities. The vehicle dynamics are given by \eqref{eq:dyn_i}. We choose velocity and turn-rate bounds as $\underline{v} = 0$ m/s, $\bar{v} = 25$ m/s, $\bar\omega = 2$ rad/s. The disturbance bound is chosen as $d_{r} = 11$ m/s, which corresponds to \textit{strong breeze} on Beaufort wind force scale \cite{Windscale}. The scheduled time of arrival $\sta$ for vehicles are chosen as $5(i-1)$ s.

The goal of the vehicles is to reach their destinations while avoiding a collision with the other vehicles. The joint state space of this 200-vehicle system is 600-dimensional, making the joint path planning and collision avoidance problem intractable for direct analysis. Therefore, we assign a priority order to vehicles and solve the path planning problem sequentially.
% !TEX root = ../../SPP_IoTjournal.tex
\subsection{Results \label{sec:bayArea_simResults}}
\begin{figure*}[!htb]
 \centering
\begin{subfigure}{0.5\textwidth}
  \includegraphics[width=\columnwidth]{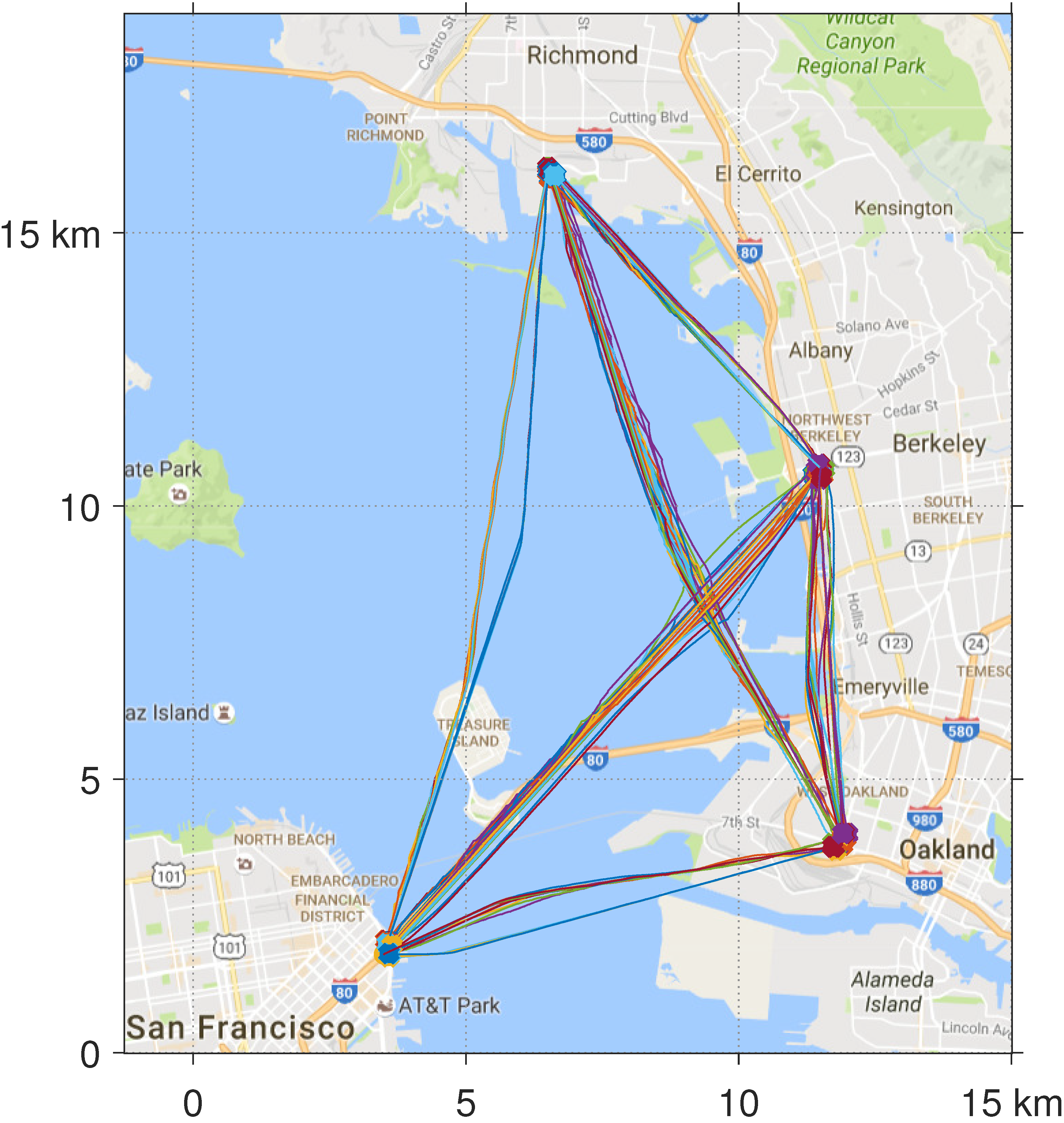}
  \subcaption{}
  \label{fig:bayArea_d11sep5}
\end{subfigure}%
\begin{subfigure}{0.5\textwidth}
  \includegraphics[width=\columnwidth]{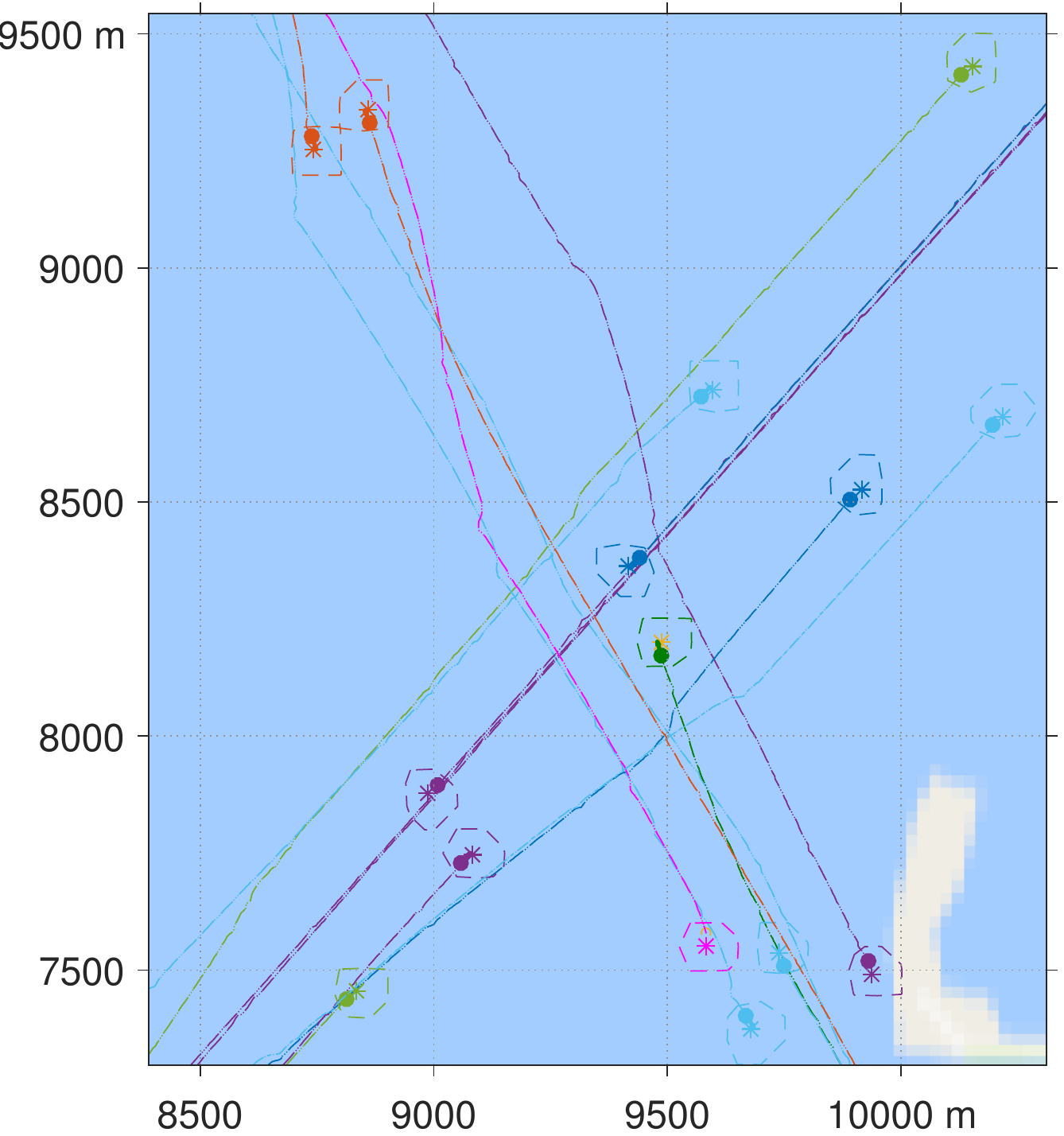}
  \subcaption{}
  \label{fig:bayArea_d11sep5_zoomed}
\end{subfigure}%
  \caption{(a) Trajectories obtained from the SPP algorithm for the multi-city simulation with $d_r = 11m/s$, $\sta_i = 5(i-1)$. (b) Zoomed-in version of the central area. A high density of vehicles is achieved at the center because of the intersection of several trajectories; however, the SPP algorithm still ensures that vehicles do not enter each other's danger zones and reach their destinations.} 
  \label{fig:bayArea_d11sep5_all}
\end{figure*}

The trajectory planning for the vehicles is done using RTT algorithm, similar to that in Section \ref{sec:city_sim}-\ref{sec:city_simResults}. The resulting trajectories of vehicles are shown in Figure \ref{fig:bayArea_d11sep5}. Once again, the vehicles remain clear of all other vehicles and reach their respective destinations. Given the separation between the scheduled times of arrival, the trajectories are predominately \textit{time-separated}, with roughly two lanes for each pair of cities (one for going from city A to city B and another for from city B to city A). A high-density of vehicles is achieved in the center since the 4 paths are intersecting in the center (Richmond-Oakland, Oakland-Richmond, Berkeley-San Francisco, San Francisco-Berkeley), but the SPP algorithm ensures safety despite this high-density, as shown in the zoomed-in version of center at an intermediate time when a large number of vehicles are passing through the central region (Figure \ref{fig:bayArea_d11sep5_zoomed}).  

\begin{figure}[t]
  \centering
  \includegraphics[width=\columnwidth]{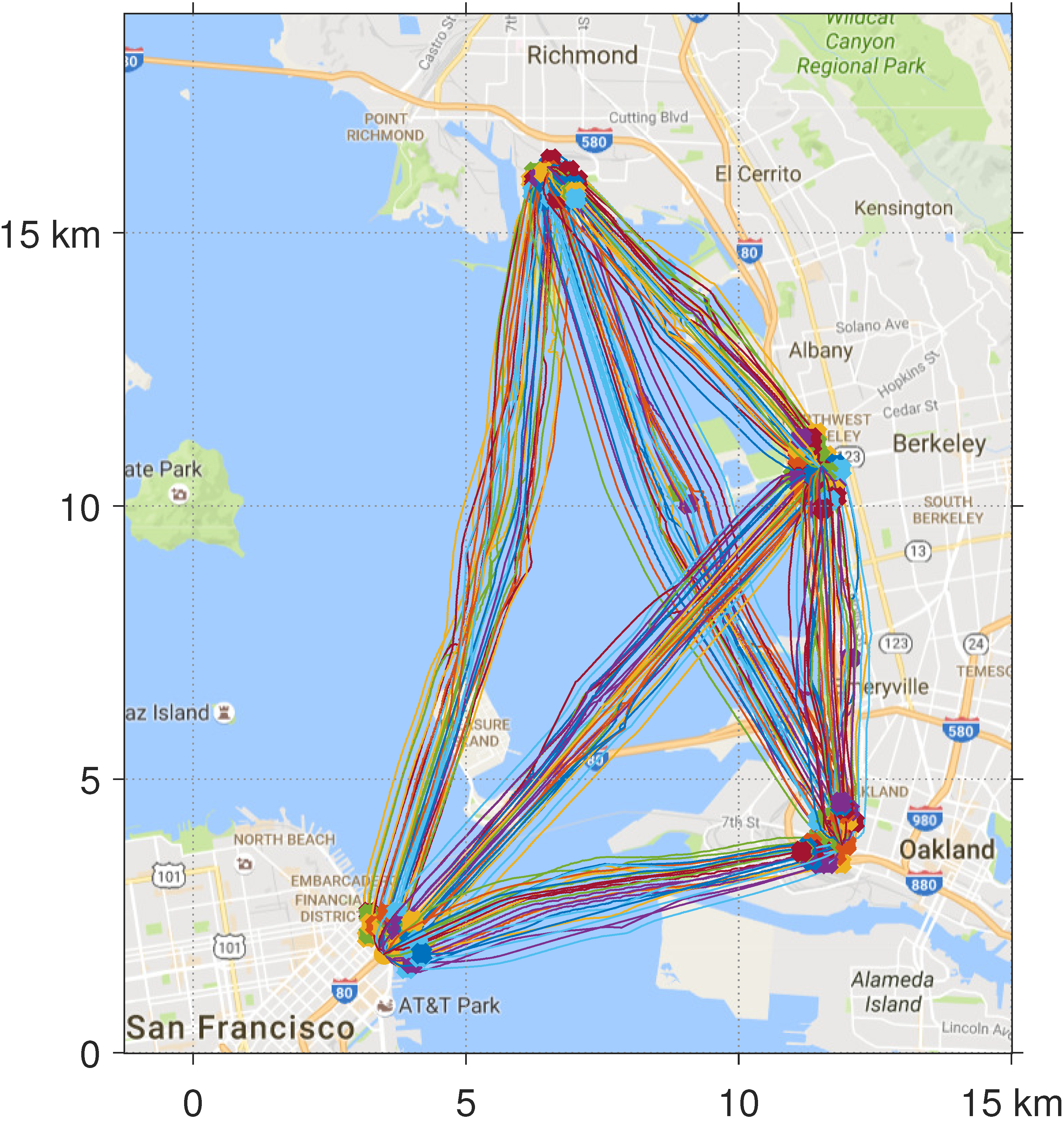}
  \caption{Vehicle trajectories for $d_r = 11m/s$, $\sta_i = 0$. Since different vehicles have same scheduled times of arrival, a multiple-lane behavior is observed between every pair of cities.} 
  \label{fig:bayArea_d11sep0}
\end{figure}

Finally, we simulate the system for the case where $\sta_i = 0 ~\forall i$. As evident from Figure \ref{fig:bayArea_d11sep0}, we get multiple lanes between each pair of cities in this case and trajectories become predominately state-separated, as we expect based on the discussion in Section \ref{sec:city_sim}-\ref{sec:city_distbEffect}.

The average computation time per vehicle is 4 minutes using a CUDA implementation of the Level Set Toolbox on a desktop computer with a Core i7 5820K processor and two GeForce GTX Titan X graphics processing units. The computation time is much longer than in the previous simulation in SF because of the larger space over which planning is done. Once again all the computation is done offline and only a lookup table query is required in real-time, which can be performed very efficiently. This simulation illustrates the scalability and the potential of deploying the SPP algorithm for provably safe path planning for large multi-vehicle systems.

% Conslusion
% !TEX root = ./SPP_IoTjournal.tex
\section{Conclusion}
Provably safe multi-vehicle path planning in an important problem that needs to be addressed to ensure that vehicles can fly in close proximity of each other. Recently, the SPP algorithm was proposed for multi-vehicle path planning problem that scales linearly with the number of vehicles. We illustrate the full potential of the algorithm by using it for large-scale multi-vehicle path planning problems under different flying conditions. We demonstrate how different types of space-time trajectories emerge naturally out of the algorithm for different disturbance conditions and other problem parameters. The reactivity of the obtained controller is also demonstrated under different wind conditions.

\section*{Acknowledgements}
This research is supported by ONR under the Embedded Humans MURI (N00014-16-1-2206).

%%%%%%%%%%%%%%%%%%%%%%%%%%%%%%%%%%%%%%%%%%%%%%%%%%%%%%%%%%%%%%%%%%%%%%%%%%%%%%%%
%\addtolength{\textheight}{1cm}   % This command serves to balance the column lengths
                                  % on the last page of the document manually. It shortens
                                  % the textheight of the last page by a suitable amount.
                                  % This command does not take effect until the next page
                                  % so it should come on the page before the last. Make
                                  % sure that you do not shorten the textheight too much.

\bibliographystyle{aiaa}
\bibliography{references}
\end{document}